\documentclass[journal]{IEEEtran}

\usepackage{bm}
\usepackage{amsopn}
\usepackage{amssymb}
\usepackage[noadjust]{cite}
\usepackage{graphicx}
\usepackage{subfigure}

\newcommand{\relvar}[2]{\buildrel {#2} \over {#1}}
\newcommand{\eqvar}[1]{\relvar{=}{\mathrm{#1}}}
\newcommand{\ltvar}[1]{\relvar{<}{\mathrm{#1}}}

\newcommand{\levar}[1]{\relvar{\le}{\mathrm{#1}}}
\newcommand{\gevar}[1]{\relvar{\ge}{\mathrm{#1}}}
\newcommand{\eqdef}{\eqvar{\scriptscriptstyle\triangle}}

\newcommand{\eqref}[1]{(\ref{#1})}

\newtheorem{theorem}{Theorem}[section]
\newtheorem{lemma}{Lemma}[section]
\newtheorem{proposition}{Proposition}[section]
\newtheorem{corollary}{Corollary}[section]
\newtheorem{remark}{Remark}[section]

\renewcommand{\IEEEQED}{\IEEEQEDopen}
\newenvironment{proofof}[1]{\begin{IEEEproof}[Proof of #1]}{\end{IEEEproof}}

\DeclareMathOperator*{\plimsup}{p-lim\,sup}
\DeclareMathOperator*{\pliminf}{p-lim\,inf}
\newcommand{\compact}{\!}

\newcommand{\seq}[1]{\mathbf{#1}}
\newcommand{\seqx}[1]{\bm{#1}}

\begin{document}

\title{Linear-Codes-Based Lossless Joint Source-Channel Coding for Multiple-Access Channels}
\author{Shengtian~Yang,~\IEEEmembership{Member,~IEEE,} Yan~Chen,~\IEEEmembership{Student Member,~IEEE,} and Peiliang~Qiu,~\IEEEmembership{Member,~IEEE}%
\thanks{This work was supported in part by the National Natural Science Foundation of China under Grants 60772093, 60802014, and 60872063, the Zhejiang Provincial Natural Science Foundation of China under Grant Y106068, and the open research fund of Key Laboratory of Information Coding and Transmission, Southwest Jiaotong University under Grant 2008-01.}
\thanks{S. Yang is with the Department of Information Science and Electronic Engineering, Zhejiang University, Hangzhou, 310027, China and also with Key Laboratory of Information Coding and Transmission, Southwest Jiaotong University, Chengdu 610031, China (e-mail: yangshengtian@zju.edu.cn).}
\thanks{Y. Chen and P. Qiu are with the Department of Information Science and Electronic Engineering, Zhejiang University, Hangzhou, 310027, China (e-mail: qiupl418@zju.edu.cn; qiupl@zju.edu.cn).}
}

\markboth{Accepted by IEEE Transactions on Information Theory (Version: \today)}{}

\pubid{0000--0000/00\$00.00~\copyright~2007 IEEE}

\maketitle

\begin{abstract}
A general lossless joint source-channel coding (JSCC) scheme based on linear codes and random interleavers for multiple-access channels (MACs) is presented and then analyzed in this paper. By the information-spectrum approach and the code-spectrum approach, it is shown that a linear code with a good joint spectrum can be used to establish limit-approaching lossless JSCC schemes for correlated general sources and general MACs, where the joint spectrum is a generalization of the input-output weight distribution. Some properties of linear codes with good joint spectra are investigated. A formula on the ``distance'' property of linear codes with good joint spectra is derived, based on which, it is further proved that, the rate of any systematic codes with good joint spectra cannot be larger than the reciprocal of the corresponding alphabet cardinality, and any sparse generator matrices cannot yield linear codes with good joint spectra. The problem of designing arbitrary rate coding schemes is also discussed. A novel idea called ``generalized puncturing'' is proposed, which makes it possible that one good low-rate linear code is enough for the design of coding schemes with multiple rates. Finally, various coding problems of MACs are reviewed in a unified framework established by the code-spectrum approach, under which, criteria and candidates of good linear codes in terms of spectrum requirements for such problems are clearly presented.
\end{abstract}


\IEEEpeerreviewmaketitle

\section{Introduction}

\IEEEPARstart{I}{n} a traditional communication system, source and channel coding are treated separately. This ascribes to the celebrated separation theorem established by Shannon in 1948, which indicates that the separation of source and channel coding incurs no loss of optimality when code length goes to infinity. While such a separate coding scheme is well motivated for the point-to-point case, it can entail significant performance losses in more general scenarios. For instance, it is well known \cite{JSCC:Cover198011} that separation between source and channel coding is not optimal in the case of transmitting correlated sources over a multiple-access channel (MAC), and the interesting toy example in \cite{JSCC:Cover198011} well illustrates the inferiority of separate coding in such cases. Moreover, even for the point-to-point case, it is recently reported by Zhong \textit{et al.} \cite{JSCC:Zhong200604} that joint source-channel coding (JSCC) usually works more efficiently (in terms of the error exponent) than separate coding. Therefore, in some applications such as the transmission of correlated sources over MACs, it is highly encouraged to adopt a JSCC scheme.

Analogous to source coding, there are two kinds of problems for JSCC, i.e., lossless JSCC and lossy JSCC. In this paper, we will only focus on lossless JSCC for MACs (also called distributed lossless JSCC). Much work on practical designs of lossless JSCC based on linear codes have been done for specific correlated sources and MACs, e.g., correlated sources over separated noisy channels \cite{JSCC:Garcia200103, JSCC:Zhong200505, JSCC:Zhao200611a}, correlated sources over additive white Gaussian noise (AWGN) MACs \cite{JSCC:Murugan200408,JSCC:Garcia200709}, and correlated sources over Rayleigh fading MACs \cite{JSCC:Zhao200611b}. However, for transmissions of arbitrary correlated sources over arbitrary MACs, it is still not clear how to construct a limit-approaching lossless JSCC scheme%
\footnote{Since there is no commonly used definition of performance limit (such as capacity) for lossless JSCC problems, the term ``limit-approaching scheme'' generally refers to the scheme that attains the best performance predicted by theory.}.
To explore the answer to this question, in this paper, we propose and analyze a general lossless JSCC scheme based on linear codes and random interleavers. The rest of this paper is organized as follows.

\IEEEpubidadjcol

In Section \ref{sec:Spectrum}, we first introduce and establish required concepts and results related to linear codes. A very important concept called spectrum is defined by the method of types \cite{JSCC:Csiszar198100}, and a series of properties about the spectrum of arbitrary mappings and linear codes are given therein.

Then, in Section \ref{sec:JSCC}, the core problem of designing limit-approaching lossless JSCC schemes is investigated. Firstly, in Section \ref{subsec:IIIb}, we show that only pairwise independence of codewords is needed in the direct part proof of JSCC, where each codeword is randomly generated according to a conditional probability distribution of channel input given source output, and the conditional probability distribution can be optimized to minimize the decoding error probability. Based on this observation, we find that a feasible method for lossless JSCC is just to find some mechanism for generating a set of pairwise independent random codewords subject to a given conditional probability distribution. Then, in Section \ref{subsec:IIIc}, by the code-spectrum approach established in Section \ref{sec:Spectrum}, we propose and analyze a lossless JSCC scheme based on linear codes and random interleavers. It is shown that linear codes with good joint spectra can be used to generate codewords satisfying pairwise independence, which then makes it possible to establish limit-approaching JSCC schemes, provided that a good conditional probability distribution is found.

In Section \ref{subsec:PropertiesOfGoodLinearCodes}, properties of linear codes with good joint spectra are investigated. In particular, a formula on the ``distance'' property of such linear codes is derived, based on which, it is further shown that the rate of any systematic codes with good joint spectra cannot be larger than the reciprocal of the corresponding alphabet cardinality, and that any sparse generator matrices cannot yield linear codes with good joint spectra. As an interesting further step, in Section \ref{subsec:VariableRateCodingSchemes}, design of arbitrary rate coding schemes is discussed, and a novel idea called "generalized puncturing" is presented therein.

All the proofs are given in Section \ref{sec:Proofs}, and Section \ref{sec:Conclusions} concludes the paper, in which we offer a complete review of various coding problems on MACs in a unified framework established by code-spectrum approach, and summarize criteria and candidates of good linear codes for each of the problems in terms of spectrum requirements.

In the sequel, symbols, real variables and deterministic mappings are denoted by lowercase letters. Sets and random elements are denoted by capital letters, and the empty set is denoted by $\emptyset$. Alphabets are all denoted by script capital letters. All logarithms are taken to the natural base $e$, and denoted by $\ln$.

\section{Linear Codes, Types, and Spectra}\label{sec:Spectrum}

In this section, we will introduce some new concepts, definitions, and notations related to linear codes. An interesting approach called ``code-spectrum'' will be established for the design and analysis of (linear) codes.

Let $\mathcal{X}$ and $\mathcal{Y}$ be two finite (additive) abelian groups, then a \emph{linear code} can be defined by a homomorphism $f: \mathcal{X}^n \to \mathcal{Y}^m$, i.e., a map satisfying
$$
f(\seq{x}_1 + \seq{x}_2) = f(\seq{x}_1) + f(\seq{x}_2), \qquad \forall \seq{x}_1, \seq{x}_2 \in \mathcal{X}^n
$$
where $\mathcal{X}^n$ and $\mathcal{Y}^m$ denote the $n$-fold direct product of $\mathcal{X}$ and the $m$-fold direct product of $\mathcal{Y}$, respectively, and $\seq{x}_1$ and $\seq{x}_2$ denote the sequences $x_{1,1}x_{1,2}\cdots x_{1,n}$ and $x_{2,1}x_{2,2}\cdots x_{2,n}$ in $\mathcal{X}^n$, respectively. We also define the rate of a linear code $f$ to be the ratio $n / m$, and denote it by $R(f)$.

A permutation on $n$ letters is defined to be a bijective map $\sigma: \mathcal{I}_n \to \mathcal{I}_n$, where
$$
\mathcal{I}_n \eqdef \{1, 2, \cdots, n\}.
$$
We denote the set of all permutations on $n$ letters by $S_n$ and the identity permutation in $S_n$ by $1_{n}$. Note that, by defining
$$
\sigma(\seq{x}) \eqdef x_{\sigma^{-1}(1)} x_{\sigma^{-1}(2)} \cdots x_{\sigma^{-1}(n)}, \qquad \forall \seq{x} \in \mathcal{X}^n
$$
any permutation (or interleaver) $\sigma$ on $n$ letters can be regarded as an automorphism on $\mathcal{X}^n$. For each positive integer $n$, we denote by $\Sigma_n$ (or $\Sigma_n'$, $\Sigma_n''$, and so on) a uniformly distributed random permutation on $n$ letters, that is
$$
\Pr\{\Sigma_n = \sigma\} = \frac{1}{|S_n|}, \qquad \forall \sigma \in S_n.
$$
We tacitly assume that different random permutations occurring in the same expression are independent, and that the notations such as $\Sigma_m$ and $\Sigma_n$ represent different random permutations though it is possible that $m = n$.

Throughout the sequel, sequences or vectors are all represented by the notations $\seq{x}$ or $(x_i)_{i \in \mathcal{I}_n}$, and its length is denoted by $|\seq{x}|$. For any part $x_ix_{i+1}\cdots x_j$ of the sequence $\seq{x}$, we use the shortening $x_{i\cdots j}$, and for $l$ times repetition of a single symbol $a$, we write $a^l$ as the shortening. When we talk about linear codes, we always assume that the domain and the range are direct products of finite abelian groups.

Next, we introduce the concept of types in the method of types \cite{JSCC:Csiszar198100}. The \emph{type} of a sequence $\seq{x}$ in $\mathcal{X}^n$ is the empirical distribution $P_{\seq{x}}$ on $\mathcal{X}$ defined by
$$
P_{\seq{x}}(a) \eqdef \frac{1}{|\seq{x}|} \sum_{i=1}^{|\seq{x}|} 1\{x_i = a\}, \qquad \forall a \in \mathcal{X}.
$$
For any distribution $P$ on $\mathcal{X}$, the set of sequences of type $P$ in $\mathcal{X}^n$ is denoted by $\mathcal{T}_P^n(\mathcal{X})$. A distribution $P$ on $\mathcal{X}$ is called a type of sequences in $\mathcal{X}^n$ if $\mathcal{T}_P^n(\mathcal{X}) \ne \emptyset$. We denote by $\mathcal{P}(\mathcal{X})$ the set of all distributions on $\mathcal{X}$, and denote by $\mathcal{P}_n(\mathcal{X})$ the set of all possible types of sequences in $\mathcal{X}^n$.

Now, we start to define the \emph{spectrum}, the most important concept in this paper. The spectrum of a set $A \subseteq \mathcal{X}^n$ is the empirical distribution $S_{\mathcal{X}}(A)$ on $\mathcal{P}_n(\mathcal{X})$ defined by
$$
S_{\mathcal{X}}(A)(P) \eqdef \frac{|\{\seq{x} \in A | P_{\seq{x}} = P\}|}{|A|}, \qquad \forall P \in \mathcal{P}_n(\mathcal{X})
$$
where $|A|$ denotes the cardinality of the set $A$. Analogously, the \emph{joint spectrum} of a set $B \subseteq \mathcal{X}^n \times \mathcal{Y}^m$ is the empirical distribution $S_{\mathcal{X}\mathcal{Y}}(B)$ on $\mathcal{P}_n(\mathcal{X}) \times \mathcal{P}_m(\mathcal{Y})$ defined by
\begin{IEEEeqnarray*}{r}
S_{\mathcal{X}\mathcal{Y}}(B)(P, Q) \eqdef \frac{|\{(\seq{x}, \seq{y}) \in B | P_{\seq{x}} = P, P_{\seq{y}} = Q\}|}{|B|} \\
\forall P \in \mathcal{P}_n(\mathcal{X}), Q \in \mathcal{P}_m(\mathcal{Y}).
\end{IEEEeqnarray*}
Furthermore, we define the \emph{marginal spectra} $S_{\mathcal{X}}(B)$ and $S_{\mathcal{Y}}(B)$ and the \emph{conditional spectra} $S_{\mathcal{Y}|\mathcal{X}}(B)$ and $S_{\mathcal{X}|\mathcal{Y}}(B)$ as the marginal distributions and the conditional distributions of $S_{\mathcal{X}\mathcal{Y}}(B)$, respectively, that is
\begin{IEEEeqnarray*}{rCl}
S_{\mathcal{X}}(B)(P) &\eqdef &\sum_{Q \in \mathcal{P}_m(\mathcal{Y})} S_{\mathcal{X}\mathcal{Y}}(B)(P, Q) \\
S_{\mathcal{Y}}(B)(Q) &\eqdef &\sum_{P \in \mathcal{P}_n(\mathcal{X})} S_{\mathcal{X}\mathcal{Y}}(B)(P, Q) \\
S_{\mathcal{Y}|\mathcal{X}}(B)(Q|P) &\eqdef &\frac{S_{\mathcal{X}\mathcal{Y}}(B)(P, Q)}{S_{\mathcal{X}}(B)(P)} \\
& &\forall P \mbox{ satisfying } S_{\mathcal{X}}(B)(P) \ne 0 \\
S_{\mathcal{X}|\mathcal{Y}}(B)(P|Q) &\eqdef &\frac{S_{\mathcal{X}\mathcal{Y}}(B)(P, Q)}{S_{\mathcal{Y}}(B)(Q)} \\
& &\forall Q \mbox{ satisfying } S_{\mathcal{Y}}(B)(Q) \ne 0.
\end{IEEEeqnarray*}

Then naturally, for any given function $f: \mathcal{X}^n \to \mathcal{Y}^m$, we may define its \emph{joint spectrum} $S_{\mathcal{X}\mathcal{Y}}(f)$, \emph{forward conditional spectrum} $S_{\mathcal{Y}|\mathcal{X}}(f)$, \emph{backward conditional spectrum} $S_{\mathcal{X}|\mathcal{Y}}(f)$, and \emph{image spectrum} $S_{\mathcal{Y}}(f)$ as $S_{\mathcal{X}\mathcal{Y}}(\mathrm{rl}(f))$, $S_{\mathcal{Y}|\mathcal{X}}(\mathrm{rl}(f))$, $S_{\mathcal{X}|\mathcal{Y}}(\mathrm{rl}(f))$, and $S_{\mathcal{Y}}(\mathrm{rl}(f))$, respectively, where $\mathrm{rl}(f)$ is the \emph{relation} defined by $\{(\seq{x}, f(\seq{x})) | \seq{x} \in \mathcal{X}^n\}$. Clearly, the forward conditional spectrum can be trivially calculated by
$$
S_{\mathcal{Y}|\mathcal{X}}(f)(Q|P) = \frac{S_{\mathcal{X}\mathcal{Y}}(f)(P, Q)}{S_{\mathcal{X}}(\mathcal{X}^n)(P)}.
$$
If $f$ is a linear code, we may further define its \emph{kernel spectrum} as $S_{\mathcal{X}}(\ker f)$, where $\ker f$ is the \emph{kernel} defined by $\{\seq{x} | f(\seq{x}) = 0^m\}$. We also have
$$
S_{\mathcal{Y}}(f) = S_{\mathcal{Y}}(f(\mathcal{X}^n))
$$
since $f$ is a homomorphism according to the definition of linear codes.

Careful readers must have noticed that the image spectrum and the joint spectrum are virtually the normalized version of the ``spectrum'' in \cite{JSCC:Bennatan200403} and the generalized and normalized version of the input-output weight distribution in \cite{JSCC:Divsalar199809}, respectively.

The above definitions can be easily extended to more general cases. For example, we may consider the joint spectrum $S_{\mathcal{X}\mathcal{Y}\mathcal{Z}}(C)$ of a set $C \subseteq \mathcal{X}^n \times \mathcal{Y}^m \times \mathcal{Z}^l$, or consider the joint spectrum $S_{\mathcal{X}_1\mathcal{X}_2\mathcal{Y}_1\mathcal{Y}_2}(g)$ of a function $g: \mathcal{X}_1^{n_1} \times \mathcal{X}_2^{n_2} \to \mathcal{Y}_1^{m_1} \times \mathcal{Y}_2^{m_2}$.

From the definitions given above, we obtain the following properties.

\begin{proposition}\label{pr:SpectrumOfSets}
For all $P \in \mathcal{P}_n(\mathcal{X})$ and $P_i \in \mathcal{P}_{n_i}(\mathcal{X}_i)$ ($1 \le i \le m$), we have
$$
S_{\mathcal{X}}(\mathcal{X}^{n})(P) = \frac{{n \choose nP}}{|\mathcal{X}|^n},
$$
$$
S_{\mathcal{X}_1 \cdots \mathcal{X}_m}(\prod_{i=1}^m A_i)(P_1, \cdots, P_m) = \prod_{i=1}^m S_{\mathcal{X}_i}(A_i)(P_i),
$$
where
$$
{n \choose nP} \eqdef \frac{n!}{\prod_{a \in \mathcal{X}} (nP(a))!}
$$
and $A_i \subseteq \mathcal{X}_i^{n_i}$ ($1 \le i \le m$).
\end{proposition}

\begin{proposition}\label{pr:RandomBin}
Let $F_{\mathcal{Y}^m | \mathcal{X}^n}^{\mathrm{RB}}: \mathcal{X}^n \to \mathcal{Y}^m$ be the \emph{random binning} function defined in \cite{JSCC:Cover197503}, that is, each $\seq{x} \in \mathcal{X}^n$ is independently assigned to one of the sequences in $\mathcal{Y}^m$ according to a uniform distribution on $\mathcal{Y}^m$, then we have
\begin{equation}\label{eq:SpectrumOfRandomBin}
E[S_{\mathcal{X}\mathcal{Y}}(F_{\mathcal{Y}^m | \mathcal{X}^n}^{\mathrm{RB}})] = S_{\mathcal{X}\mathcal{Y}}(\mathcal{X}^n \times \mathcal{Y}^m).
\end{equation}
Generally, we have
\begin{IEEEeqnarray}{Cl}
&E[S_{\mathcal{X}_1\cdots\mathcal{X}_l \mathcal{Y}_1\cdots\mathcal{Y}_k}(F_{\prod_{i=1}^k \mathcal{Y}_i^{m_i} | \prod_{i=1}^l \mathcal{X}_i^{n_i}}^{\mathrm{RB}})] \IEEEnonumber\\
= &S_{\mathcal{X}_1\cdots\mathcal{X}_l \mathcal{Y}_1\cdots\mathcal{Y}_k}(\prod_{i=1}^l \mathcal{X}_i^{n_i} \times \prod_{i=1}^k \mathcal{Y}_i^{m_i}). \label{eq:Spectrum2OfRandomBin}
\end{IEEEeqnarray}
\end{proposition}

\begin{proposition}\label{pr:SpectrumInvarianceUnderPermutation}
For any given function $f: \prod_{i=1}^l \mathcal{X}_i^{n_i} \to \prod_{i=1}^k \mathcal{Y}_i^{m_i}$ and any permutations $\{\sigma_i \in S_{n_i}\}_{i=1}^l$ and $\{\sigma_i' \in S_{m_i}\}_{i=1}^k$, we have
$$
S_{\mathcal{X}_1\cdots\mathcal{X}_l \mathcal{Y}_1\cdots\mathcal{Y}_k}(\seqx{\sigma}' \circ f \circ \seqx{\sigma})
= S_{\mathcal{X}_1\cdots\mathcal{X}_l \mathcal{Y}_1\cdots\mathcal{Y}_k}(f),
$$
where $(f \circ g)(x) \eqdef f(g(x))$ and
$$
\seqx{\sigma}((\seq{x}_i)_{i \in \mathcal{I}_l}) \eqdef (\sigma_{i}(\seq{x}_i))_{i \in \mathcal{I}_l}
$$
$$
\seqx{\sigma}'((\seq{y}_i)_{i \in \mathcal{I}_k}) \eqdef (\sigma_{i}'(\seq{y}_i))_{i \in \mathcal{I}_k}.
$$
\end{proposition}

\begin{proposition}\label{pr:SpectrumPropertyOfFunctions}
For any given random function $F: \mathcal{X}^n \to \mathcal{Y}^m$, we have
\begin{equation}\label{eq:SpectrumPropertyOfFunctions}
\Pr\{\tilde{F}(\seq{x}) = \seq{y}\} = |\mathcal{Y}|^{-m} \alpha(F)(P_{\seq{x}}, P_{\seq{y}})
\end{equation}
for any $\seq{x} \in \mathcal{X}^n$ and $\seq{y} \in \mathcal{Y}^m$, where
\begin{equation}\label{eq:RandomizedF1}
\tilde{F} \eqdef \Sigma_m \circ F \circ \Sigma_n
\end{equation}
and
\begin{IEEEeqnarray}{rCl}
\alpha(F)(P, Q) &\eqdef &\frac{E[S_{\mathcal{X}\mathcal{Y}}(F)(P, Q)]}{S_{\mathcal{X}\mathcal{Y}}(\mathcal{X}^n \times \mathcal{Y}^m)(P, Q)} \IEEEnonumber \\
&= &\frac{E[S_{\mathcal{Y}|\mathcal{X}}(F)(Q|P)]}{S_{\mathcal{Y}}(\mathcal{Y}^m)(Q)}. \label{eq:DefinitionOfAlpha}
\end{IEEEeqnarray}
\end{proposition}

\begin{proposition}\label{pr:GoodLinearCodes}
For any given linear code $f: \mathcal{X}^n \to \mathcal{Y}^m$, we have
\begin{equation}\label{eq:Identity1OfLC}
\Pr\{f(0^n) = \seq{y}\} = 1\{\seq{y} = 0^m\}
\end{equation}
or equivalently
\begin{equation}\label{eq:Identity2OfLC}
\alpha(f)(P_{0^n}, P_{\seq{y}}) = |\mathcal{Y}|^{m} 1\{\seq{y} = 0^m\}.
\end{equation}
If both $\mathcal{X}$ and $\mathcal{Y}$ are the finite field $\mathbb{F}_q$, we define a random linear code $F_{q, n, m}^{\mathrm{RLC}}: \mathcal{X}^n \to \mathcal{Y}^m$ given by $\seq{x} \mapsto \seq{x} A_{n \times m}$, where $\seq{x}$ represents an $n$-dimensional row vector, and $A_{n \times m}$ denotes a random $n \times m$ matrix with each entry independently taking values in $\mathbb{F}_q$ according to a uniform distribution on $\mathbb{F}_q$. (Note that for each realization of $A_{n \times m}$, we then obtain a corresponding realization of $F_{q, n, m}^{\mathrm{RLC}}$, and such a random construction has ever been adopted in \cite[Section 2.1]{JSCC:Gallager196300}, \cite{JSCC:Csiszar198207}, etc.) Then, we have
\begin{equation}\label{eq:Identity1OfGLC}
\Pr\{F_{q, n, m}^{\mathrm{RLC}}(\seq{x}) = \seq{y}\} = q^{-m}
\end{equation}
for all $\seq{x} \in \mathcal{X}^n \backslash \{0^n\}$ and $\seq{y} \in \mathcal{Y}^m$, or equivalently
\begin{equation}\label{eq:Identity2OfGLC}
\alpha(F_{q, n, m}^{\mathrm{RLC}})(P, Q) = 1
\end{equation}
for all $P \in \mathcal{P}_n(\mathcal{X}) \backslash \{P_{0^n}\}$ and $Q \in \mathcal{P}_m(\mathcal{Y})$.
\end{proposition}

\begin{proposition}\label{pr:IndependenceLC1}
For a given random linear code $F: \mathcal{X}^n \to \mathcal{Y}^m$, we have
\begin{IEEEeqnarray}{l}
\Pr\{\hat{F}(\seq{x}) = \seq{y}\} = |\mathcal{Y}|^{-m} \label{eq:Identity1OfIndependenceLC1}\\
\Pr\{\hat{F}(\hat{\seq{x}}) = \hat{\seq{y}} | \hat{F}(\seq{x}) = \seq{y}\} \IEEEnonumber\\
\qquad = |\mathcal{Y}|^{-m} \alpha(F)(P_{\hat{\seq{x}}-\seq{x}}, P_{\hat{\seq{y}}-\seq{y}}) \label{eq:Identity2OfIndependenceLC1}
\end{IEEEeqnarray}
for any unequal $\seq{x}, \hat{\seq{x}} \in \mathcal{X}^n$ and any $\seq{y}, \hat{\seq{y}} \in \mathcal{Y}^m$, where
\begin{equation}\label{eq:RandomizedF2}
\hat{F}(\seq{x}) \eqdef \tilde{F}(\seq{x}) + \bar{Y}^m, \qquad \forall \seq{x} \in \mathcal{X}^n
\end{equation}
and $\bar{Y}^m$ denotes an independent uniform random vector on $\mathcal{Y}^m$.
\end{proposition}

The proofs of Proposition \ref{pr:SpectrumOfSets} and \ref{pr:SpectrumInvarianceUnderPermutation} are easy and hence omitted in this paper, and the proofs of the other propositions are presented in Section \ref{subsec:ProofOfSectionII}.

\section{Lossless Joint Source-Channel Coding Based on Linear Codes and Random Interleavers}\label{sec:JSCC}

In this section, a lossless JSCC scheme based on linear codes will be presented for the transmission of arbitrary correlated sources over arbitrary MACs, and an information-spectrum approach (\cite[pp. 247-268]{JSCC:Han200300}, \cite{JSCC:Iwata200511}) is adopted to analyze the performance of the scheme.

\subsection{Preliminaries of Information-Spectrum and Formulation of the Lossless JSCC problem}

At first, we introduce some concepts of limits in probability, which are frequently used in the methods of information spectrum \cite{JSCC:Han200300}. For a sequence $\{Z_n\}_{n=1}^\infty$ of real-valued random variables, the \emph{limit superior in probability} and the \emph{limit inferior in probability} of $\{Z_n\}_{n=1}^\infty$ are defined by
\begin{equation}\label{eq:DefinitionOfPLimsup}
\plimsup_{n \to \infty} Z_n \eqdef \inf\biggl\{\alpha \bigg| \lim_{n \to \infty} \Pr\{Z_n > \alpha\} = 0 \biggr\}
\end{equation}
and
\begin{equation}\label{eq:DefinitionOfPLiminf}
\pliminf_{n \to \infty} Z_n \eqdef \sup\biggl\{\beta \bigg| \lim_{n \to \infty} \Pr\{Z_n < \beta\} = 0 \biggr\}
\end{equation}
respectively.

Secondly, in the methods of information spectrum, a \emph{general source} is defined as an infinite sequence
$$
\bm{X} = \{X^n = (X_1^{(n)}, X_2^{(n)}, \cdots, X_n^{(n)})\}_{n=1}^{\infty}
$$
of $n$-dimensional random variables $X^n$ where each component random variable $X_i^{(n)}$ ($1 \le i \le n$) takes values in the alphabet $\mathcal{X}$, and a \emph{general channel} is defined as an infinite sequence $\bm{W} = \{W^n\}_{n=1}^{\infty}$ of conditional probability distributions $W^n: \mathcal{X}^n \to \mathcal{Y}^n$ satisfying
$$
\sum_{\seq{y} \in \mathcal{Y}^n} W^n(\seq{y}|\seq{x}) = 1, \qquad \forall \seq{x} \in \mathcal{X}^n
$$
for each $n = 1, 2, \ldots$. Analogously, a $K$-tuple of correlated general sources are defined to be an infinite sequence
$$
(\bm{V}_i)_{i \in \mathcal{I}_K} \eqdef \{(V_i^n)_{i \in \mathcal{I}_K}\}_{n=1}^{\infty}
$$
of random vectors $(V_i^n)_{i \in \mathcal{I}_K}$ taking values in $\prod_{i \in \mathcal{I}_K} \mathcal{V}_i^n$. Also, a general multiple-access channel (MAC) with $K$ input terminals is defined to be an infinite sequence $\bm{W} = \{W^n\}_{n=1}^{\infty}$ of conditional probability distributions $W^n: \prod_{i \in \mathcal{I}_K} \mathcal{X}_i^n \to \mathcal{Y}^n$ satisfying
$$
\sum_{\seq{y} \in \mathcal{Y}^n} W^n(\seq{y}|(\seq{x}_i)_{i \in \mathcal{I}_K}) = 1, \qquad \forall (\seq{x}_i)_{i \in \mathcal{I}_K} \in \prod_{i \in \mathcal{I}_K} \mathcal{X}_i^n
$$
for each $n = 1, 2, \ldots$. To simplify notations, for any nonempty set $A \subseteq \mathcal{I}_K$, we further define
\begin{IEEEeqnarray*}{rCl}
\bm{X}_A &\eqdef &(\bm{X}_i)_{i \in A} \\
X_A^n &\eqdef &(X_i^n)_{i \in A} \\
\seq{x}_A &\eqdef &(\seq{x}_i)_{i \in A} \in \prod_{i \in A} \mathcal{X}_i^n.
\end{IEEEeqnarray*}

Next, for any given general sources $\bm{X}$ and $\bm{Y}$, the \emph{spectral sup-entropy rate} of $\bm{X}$ and the \emph{spectral inf-mutual information rate} between $\bm{X}$ and $\bm{Y}$ are defined by
\begin{equation}
\overline{H}(\bm{X}) \eqdef \plimsup_{n \to \infty} \frac{1}{n} \ln \frac{1}{P_{X^n}(X^n)}
\end{equation}
and
\begin{equation}
\underline{I}(\bm{X}; \bm{Y}) \eqdef \pliminf_{n \to \infty} \frac{1}{n} \ln \frac{P_{X^n|Y^n}(X^n|Y^n)}{P_{X^n}(X^n)}
\end{equation}
respectively. Moreover, for any given general sources $\bm{X}$, $\bm{Y}$, and $\bm{Z}$, the \emph{spectral conditional sup-entropy rate} of $\bm{X}$ given $\bm{Z}$ and the \emph{spectral conditional inf-mutual information rate} between $\bm{X}$ and $\bm{Y}$ given $\bm{Z}$ are defined by
\begin{equation}
\overline{H}(\bm{X}|\bm{Z}) \eqdef \plimsup_{n \to \infty} \frac{1}{n} \ln \frac{1}{P_{X^n|Z^n}(X^n|Z^n)}
\end{equation}
and
\begin{equation}
\underline{I}(\bm{X}; \bm{Y}|\bm{Z}) \eqdef \pliminf_{n \to \infty} \frac{1}{n} \ln \frac{P_{X^n|Y^nZ^n}(X^n|Y^n,Z^n)}{P_{X^n|Z^n}(X^n|Z^n)}
\end{equation}
respectively.

Finally, based on the notations above, a lossless JSCC problem for correlated sources and MACs may be described as follows.

Suppose that a $K$-tuple of correlated general sources $\bm{V}_{\mathcal{I}_K}$ and an arbitrary general MAC $\bm{W}$ with $K$ input terminals are given, we consider $K$ sequences of separate \emph{encoders} $\varphi_n^{(i)}: \mathcal{V}_i^n \to \mathcal{X}_i^n$ ($i \in \mathcal{I}_K$) and one sequence of joint \emph{decoders} $\psi_n: \mathcal{Y}^n \to \prod_{i \in \mathcal{I}_K} \mathcal{V}_i^n$, where $\psi_n$ may also be expressed as $(\psi_n^{(i)})_{i \in \mathcal{I}_K}$. The whole coding system then works in the following way.

Each encoder $\varphi_n^{(i)}$ ($i \in \mathcal{I}_K$) separately encodes the corresponding $n$-length source output $V_i^n$ into a codeword $X_i^n = \varphi_n^{(i)}(V_i^n)$, and then the codewords $X_{\mathcal{I}_{K}}^n$ are transmitted over the general MAC $W^n$ simultaneously. The joint decoder $\psi_n$ observes the output $Y^n$ from the MAC $W^n$ to reproduce the estimates $\hat{V}_{\mathcal{I}_K}^n \eqdef \psi_n(Y^n) = (\psi_n^{(i)}(Y^n))_{i \in \mathcal{I}_K}$ of $V_{\mathcal{I}_K}^n$.

To evaluate the performance of such a system, we define the \emph{decoding error probability} $\epsilon_n$ as the average decoding error probability with respect to the probability distribution $P_{V_{\mathcal{I}_K}^n}$ of the correlated sources, i.e.,
\begin{IEEEeqnarray*}{rCl}
\epsilon_n &\eqdef &\Pr\{V_{\mathcal{I}_K}^n \ne \psi_n(Y^n)\} \\
&= &\sum_{\seq{v}_{\mathcal{I}_K} \in \prod_{i \in \mathcal{I}_K} \mathcal{V}_i^n} P_{V_{\mathcal{I}_K}^n}(\seq{v}_{\mathcal{I}_K}) \\
& &W^n([\psi_n^{-1}(\seq{v}_{\mathcal{I}_K})]^c | (\varphi_n^{(i)}(\seq{v}_i))_{i \in \mathcal{I}_K}).
\end{IEEEeqnarray*}
For simplicity, we call a tuple $((\varphi_n^{(i)})_{i \in \mathcal{I}_K}, \psi_n)$ of $K$ encoders and one decoder with the decoding error probability $\epsilon_n$ an $(n, \epsilon_n)$-code. Then, for a $K$-tuple of correlated general sources $\bm{V}_{\mathcal{I}_K}$ and a general MAC $\bm{W}$, we define the sources $\bm{V}_{\mathcal{I}_K}$ being $\epsilon$-transmissible over the channel $\bm{W}$ if there exists an $(n, \epsilon_n)$-code satisfying
$$
\limsup_{n \to \infty} \epsilon_n \le \epsilon.
$$
When $\epsilon = 0$, we say that $\bm{V}_{\mathcal{I}_K}$ is transmissible over channel $\bm{W}$.

\subsection{Essential Properties for Designs of JSCC Schemes}\label{subsec:IIIb}

Now, we start to consider the design of a lossless JSCC scheme based on linear codes. Before presenting our scheme, we first prove some results to show some essential properties, which are helpful for establishing a limit-approaching scheme.

\begin{lemma}\label{le:FeinsteinLemmaForMAC}
Let $\bm{V}_{\mathcal{I}_K}$ be a $K$-tuple of correlated general sources and $\bm{W}$ a MAC with $K$ input terminals. For a given tuple $(\Phi_n^{(i)})_{i \in \mathcal{I}_K}$ of random encoders, if
\begin{IEEEeqnarray}{l}
\Pr\{(\Phi_n^{(i)}(\seq{v}_i))_{i \in \mathcal{I}_K} = \seq{x}_{\mathcal{I}_K}\} = \prod_{i \in \mathcal{I}_K} \Pr\{\Phi_n^{(i)}(\seq{v}_i) = \seq{x}_i\} \IEEEnonumber \\
\hspace{7em} \forall \seq{v}_{\mathcal{I}_K} \in \prod_{i \in \mathcal{I}_K} \mathcal{V}_i^n, \seq{x}_{\mathcal{I}_K} \in \prod_{i \in \mathcal{I}_K} \mathcal{X}_i^n \label{eq:Condition1OfEncoders}
\end{IEEEeqnarray}
and there exits a family of real functions
$$
\bigl\{ \rho(\Phi_n^{(A)}): \prod_{i \in A} \mathcal{V}_i^n \times \prod_{i \in A} \mathcal{X}_i^n \to \mathbb{R} \bigr\}_{A \subseteq \mathcal{I}_K, A \ne \emptyset}
$$
satisfying
\begin{IEEEeqnarray}{Cl}
&\Pr\{(\Phi_n^{(i)}(\hat{\seq{v}}_i))_{i \in A} = \hat{\seq{x}}_A | (\Phi_n^{(i)}(\seq{v}_i))_{i \in \mathcal{I}_K} = \seq{x}_{\mathcal{I}_K}\} \IEEEnonumber \\
\le &e^{n\rho(\Phi_n^{(A)})(\hat{\seq{v}}_A, \hat{\seq{x}}_A)} \prod_{i \in A} \Pr\{\Phi_n^{(i)}(\hat{\seq{v}}_i) = \hat{\seq{x}}_i\} \label{eq:Condition2OfEncoders}
\end{IEEEeqnarray}
for all nonempty set $A \subseteq \mathcal{I}_K$, any unequal $\seq{v}_i, \hat{\seq{v}}_i \in \mathcal{V}_i^n$ ($i \in \mathcal{I}_K$), and any $\seq{x}_i, \hat{\seq{x}}_i \in \mathcal{X}_i^n$ ($i \in \mathcal{I}_K$), then there exists a joint decoder $\Psi_n$ such that the decoding error probability
\begin{equation}\label{eq:UBofEpsilon}
\epsilon_n \le \Pr\left\{ (V_{\mathcal{I}_K}^n, X_{\mathcal{I}_K}^n, Y^n) \not \in T_{n,\gamma}(\Phi_n) \right\} + (2^K - 1)e^{-n\gamma}
\end{equation}
for all $n = 1, 2, \ldots$, where $\gamma > 0$ and
\begin{IEEEeqnarray}{l}
P_{V_{\mathcal{I}_K}^nX_{\mathcal{I}_K}^nY^n}(\seq{v}_{\mathcal{I}_K}, \seq{x}_{\mathcal{I}_K}, \seq{y}) = \IEEEnonumber \\
\qquad P_{V_{\mathcal{I}_K}^n}(\seq{v}_{\mathcal{I}_K}) W^n(\seq{y} | \seq{x}_{\mathcal{I}_K}) \prod_{i \in \mathcal{I}_K} P_{X_i^n|V_i^n}(\seq{x}_i | \seq{v}_i) \label{eq:DistributionOfVXY} \\
P_{X_i^n|V_i^n}(\seq{x}_i | \seq{v}_i) = \Pr\{\Phi_n^{(i)}(\seq{v}_i) = \seq{x}_i\} \label{eq:ConditionalDistributionX|V}
\end{IEEEeqnarray}
\begin{IEEEeqnarray}{rCl}
T_{n,\gamma}(\Phi_n) &= &\bigcap_{A \subseteq \mathcal{I}_K, A \ne \emptyset} T_{n,\gamma}^A(\Phi_n) \label{eq:GoodSetForDecoding} \\
T_{n,\gamma}^A(\Phi_n) &= &\biggl\{ (\seq{v}_{\mathcal{I}_K}, \seq{x}_{\mathcal{I}_K}, \seq{y}) \in \prod_{i \in \mathcal{I}_K} \mathcal{V}_i^n \times \prod_{i \in \mathcal{I}_K} \mathcal{X}_i^n \times \mathcal{Y}^n \bigg| \IEEEnonumber \\
& &i(\seq{x}_A; \seq{y} | \seq{x}_{A^c}, \seq{v}_{A^c}) > h(\seq{v}_A | \seq{v}_{A^c}) \IEEEnonumber \\
& &+\: \rho(\Phi_n^{(A)})(\seq{v}_A, \seq{x}_A) + \gamma \biggr\} \label{eq:GoodSet2ForDecoding}
\end{IEEEeqnarray}
and
\begin{IEEEeqnarray*}{rCl}
A^c &\eqdef &\mathcal{I}_K \backslash A \\
i(\seq{x}_A; \seq{y}|\seq{x}_{A^c}, \seq{v}_{A^c}) &\eqdef &\frac{1}{n} \ln \frac{P_{Y^n|X_{\mathcal{I}_K}^nV_{A^c}^n}(\seq{y}|\seq{x}_{\mathcal{I}_K}, \seq{v}_{A^c})}{P_{Y^n|X_{A^c}^nV_{A^c}^n}(\seq{y}|\seq{x}_{A^c}, \seq{v}_{A^c})} \IEEEnonumber\\
&= &\frac{1}{n} \ln \frac{W^n(\seq{y}|\seq{x}_{\mathcal{I}_K})}{P_{Y^n|X_{A^c}^nV_{A^c}^n}(\seq{y}|\seq{x}_{A^c}, \seq{v}_{A^c})} \\
h(\seq{v}_A|\seq{v}_{A^c}) &\eqdef &\frac{1}{n} \ln \frac{1}{P_{V_A^n|V_{A^c}^n}(\seq{v}_A|\seq{v}_{A^c})}.
\end{IEEEeqnarray*}
\end{lemma}

\begin{remark}\label{re:1stForFeinsteinLemmaForMAC}
At first, note that the requirement \eqref{eq:Condition2OfEncoders} is only valid for $\Pr\{(\Phi_n^{(i)}(\seq{v}_i))_{i \in \mathcal{I}_K} = \seq{x}_{\mathcal{I}_K}\} > 0$. As for the case of $\Pr\{(\Phi_n^{(i)}(\seq{v}_i))_{i \in \mathcal{I}_K} = \seq{x}_{\mathcal{I}_K}\} = 0$, we need to consider the alternative form of \eqref{eq:Condition2OfEncoders}, that is
\begin{IEEEeqnarray*}{Cl}
&\Pr\{(\Phi_n^{(i)}(\hat{\seq{v}}_i))_{i \in A} = \hat{\seq{x}}_A, (\Phi_n^{(i)}(\seq{v}_i))_{i \in \mathcal{I}_K} = \seq{x}_{\mathcal{I}_K}\} \\
\le &e^{n\rho(\Phi_n^{(A)})(\hat{\seq{v}}_A, \hat{\seq{x}}_A)} \prod_{i \in A} \Pr\{\Phi_n^{(i)}(\hat{\seq{v}}_i) = \hat{\seq{x}}_i\} \\
&\Pr\{(\Phi_n^{(i)}(\seq{v}_i))_{i \in \mathcal{I}_K} = \seq{x}_{\mathcal{I}_K}\}.
\end{IEEEeqnarray*}
Clearly, for $\Pr\{(\Phi_n^{(i)}(\seq{v}_i))_{i \in \mathcal{I}_K} = \seq{x}_{\mathcal{I}_K}\} = 0$, the above inequality holds for any $\rho(\Phi_n^{(A)})(\hat{\seq{v}}_A, \hat{\seq{x}}_A) < \infty$. So in the sequel, this singular case will be ignored for simplicity, and such a simplification has no impact on the correctness of the results. Next, for any tuple $(\Phi_n^{(i)})_{i \in \mathcal{I}_K}$ of encoders, we can define
\begin{IEEEeqnarray*}{l}
\rho(\Phi_n^{(A)})(\hat{\seq{v}}_A, \hat{\seq{x}}_A) \eqdef \max_{\scriptstyle \seq{v}_{\mathcal{I}_K} \in \prod_{i \in \mathcal{I}_K} \mathcal{V}_i^n, (\seq{v}_i \ne \hat{\seq{v}}_i)_{i \in A}, \atop \scriptstyle \seq{x}_{\mathcal{I}_K} \in \prod_{i \in \mathcal{I}_K} \mathcal{X}_i^n} \\
\quad \frac{1}{n} \ln \frac{\Pr\{(\Phi_n^{(i)}(\hat{\seq{v}}_i))_{i \in A} = \hat{\seq{x}}_A | (\Phi_n^{(i)}(\seq{v}_i))_{i \in \mathcal{I}_K} = \seq{x}_{\mathcal{I}_K}\}}{\prod_{i \in A} \Pr\{\Phi_n^{(i)}(\hat{\seq{v}}_i) = \hat{\seq{x}}_i\}}
\end{IEEEeqnarray*}
for all $(\hat{\seq{v}}_A, \hat{\seq{x}}_A)$ satisfying $\Pr\{\Phi_n^{(i)}(\hat{\seq{v}}_i) = \hat{\seq{x}}_i\} \ne 0$ for each $i \in A$ and
$$
\rho(\Phi_n^{(A)})(\hat{\seq{v}}_A, \hat{\seq{x}}_A) = 0
$$
for other $(\hat{\seq{v}}_A, \hat{\seq{x}}_A)$, so there always exist well-defined functions $\{\rho(\Phi_n^{(A)})\}_{A \subseteq \mathcal{I}_K, A \ne \emptyset}$. If the tuple $(\Phi_n^{(A)})_{i \in \mathcal{I}_K}$ of encoders are all independent, then the condition \eqref{eq:Condition2OfEncoders} may be reduced to the following one:
\begin{IEEEeqnarray}{l}
\Pr\{\Phi_n^{(i)}(\hat{\seq{v}}_i) = \hat{\seq{x}}_i | \Phi_n^{(i)}(\seq{v}_i) = \seq{x}_i\} \IEEEnonumber \\
\quad \le e^{n\rho(\Phi_n^{(i)})(\hat{\seq{v}}_i, \hat{\seq{x}}_i)} \Pr\{\Phi_n^{(i)}(\hat{\seq{v}}_i) = \hat{\seq{x}}_i\} \label{eq:Condition3OfEncoders}
\end{IEEEeqnarray}
for any unequal $\seq{v}_i, \hat{\seq{v}}_i \in \mathcal{V}_i^n$ and any $\seq{x}_i, \hat{\seq{x}}_i \in \mathcal{X}_i^n$ ($i \in \mathcal{I}_K$), and we have
\begin{equation}
\rho(\Phi_n^{(A)})(\seq{v}_A, \seq{x}_A) = \sum_{i \in A} \rho(\Phi_n^{(i)})(\seq{v}_i, \seq{x}_i)
\end{equation}
for any nonempty set $A \subseteq \mathcal{I}_K$. Furthermore, if we define each encoder $\Phi_n^{(i)}$ by independently generate each encoder output $\Phi_n^{(i)}(\seq{v}_i)$ according to a fixed $P_{\mathcal{X}_i^n|\mathcal{V}_i^n}(\cdot|\seq{v}_i)$, then we have
\begin{equation}\label{eq:PairwiseIndependence}
\Pr\{\Phi_n^{(i)}(\hat{\seq{v}}_i) = \hat{\seq{x}}_i | \Phi_n^{(i)}(\seq{v}_i) = \seq{x}_i\} \equiv \Pr\{\Phi_n^{(i)}(\hat{\seq{v}}_i) = \hat{\seq{x}}_i\}.
\end{equation}
Therefore
\begin{equation}\label{eq:RandomCodeForMAC}
\rho(\Phi_n^{(A)}) \equiv 0, \qquad \forall A \in \mathcal{I}_K, A \ne \emptyset.
\end{equation}
The above construction has been adopted in the proofs of \cite[Lemma 3.8.1]{JSCC:Han200300} and \cite[Lemma 1]{JSCC:Iwata200511}. We denote this special construction by $(\mathcal{F}_n^{(i)})_{i \in \mathcal{I}_k}$. The significance and implications of \eqref{eq:RandomCodeForMAC} will be discussed later in Remark \ref{re:ArtOfJSCC}, but readers should be reminded here that the construction $(\mathcal{F}_n^{(i)})_{i \in \mathcal{I}_k}$ is only one of the methods for constructing encoders satisfying \eqref{eq:RandomCodeForMAC}, since in \eqref{eq:PairwiseIndependence}, only pairwise independence instead of mutual independence are required to yield \eqref{eq:RandomCodeForMAC}.
\end{remark}

The proof of Lemma \ref{le:FeinsteinLemmaForMAC} is presented in Section \ref{subsec:ProofOfSectionIII}, and it is an easy extension of the proof of Lemma 1 in \cite{JSCC:Iwata200511}. By Lemma \ref{le:FeinsteinLemmaForMAC}, we immediately obtain the following theorem.

\begin{theorem}\label{th:GeneralizedDirectPartForMAC}
Let $\bm{V}_{\mathcal{I}_K}$ be a $K$-tuple of correlated general sources, $\bm{W}$ a general MAC with $K$ input terminals, and $(\Phi_n^{(i)})_{i \in \mathcal{I}_K}$ a given tuple of random encoders satisfying \eqref{eq:Condition1OfEncoders} and \eqref{eq:Condition2OfEncoders}. If for a  sequence $\{\gamma_n\}_{n=1}^\infty$ satisfying
$$
\gamma_n > 0, \gamma_n \to 0, \mbox{ and } n\gamma_n \to \infty, \qquad \mbox{as } n \to \infty
$$
it holds that
\begin{equation}\label{eq:ConditionOfGeneralizedDirectPartForMAC}
\limsup_{n \to \infty} \Pr\left\{ (V_{\mathcal{I}_K}^n, X_{\mathcal{I}_K}^n, Y^n) \not \in T_{n,\gamma_n}(\Phi_n) \right\} \le \epsilon
\end{equation}
then there exists an $(n, \epsilon_n)$ code based on $(\Phi_n^{(i)})_{i \in \mathcal{I}_K}$ such that $\bm{V}_{\mathcal{I}_K}$ is $\epsilon$-transmissible over $\bm{W}$, where the probability distribution of $(V_{\mathcal{I}_K}^n, X_{\mathcal{I}_K}^n, Y^n)$ is defined by \eqref{eq:DistributionOfVXY}, and $T_{n,\gamma_n}(\Phi_n)$ is defined by \eqref{eq:GoodSetForDecoding}.
\end{theorem}

Theorem \ref{th:GeneralizedDirectPartForMAC} gives a way for estimating the upper bound of the optimal decoding error probability of arbitrary encoding schemes for arbitrary sources and MACs. From Theorem \ref{th:GeneralizedDirectPartForMAC}, a useful corollary follows.

\begin{corollary}\label{co:GeneralizedDirectPartForMAC}
Let $\bm{V}_{\mathcal{I}_K}$ be a $K$-tuple of correlated general sources, $\bm{W}$ a general MAC with $K$ input terminals, and $(\Phi_n^{(i)})_{i \in \mathcal{I}_K}$ a given tuple of random encoders satisfying \eqref{eq:Condition1OfEncoders} and \eqref{eq:Condition2OfEncoders}. If it holds that
\begin{IEEEeqnarray}{rCl}
\overline{H}(\bm{V}_A|\bm{V}_{A^c}) &< &\underline{I}(\bm{X}_A; \bm{Y}|\bm{X}_{A^c}, \bm{V}_{A^c}) \IEEEnonumber \\
& &-\: \plimsup_{n \to \infty} \rho(\Phi_n^{(A)})(V_A^n, X_A^n) \label{eq:AchievableCondition}
\end{IEEEeqnarray}
for all nonempty set $A \subseteq \mathcal{I}_K$, then there exists an $(n, \epsilon_n)$ code based on $(\Phi_n^{(i)})_{i \in \mathcal{I}_K}$ such that $\bm{V}_{\mathcal{I}_K}$ is transmissible over $\bm{W}$, where the probability distribution of $(V_{\mathcal{I}_K}^n, X_{\mathcal{I}_K}^n, Y^n)$ is defined by \eqref{eq:DistributionOfVXY}.
\end{corollary}

The proof of Corollary \ref{co:GeneralizedDirectPartForMAC} is presented in Section \ref{subsec:ProofOfSectionIII}.

Furthermore, assuming that the sources and channels are stationary and memoryless, we prove the following corollary. For convenience, a $K$-tuple of correlated stationary memoryless sources are now represented by a random vector $V_{\mathcal{I}_K}$, and a stationary memoryless MAC with $K$ input terminals is represented by a conditional probability distribution $W$ satisfying $W^n(\seq{y}|\seq{x}_{\mathcal{I}_K}) = \prod_{i=1}^n W(y_i|x_{\mathcal{I}_K,i})$, where $x_{\mathcal{I}_K,i} \eqdef x_{1\cdots K, i}$.

\begin{corollary}\label{co:DirectPartForMemorylessMAC}
Let $V_{\mathcal{I}_K}$ be a $K$-tuple of correlated stationary memoryless sources, $W$ a stationary memoryless MAC with $K$ input terminals, and $(\Phi_n^{(i)})_{i \in \mathcal{I}_K}$ a given tuple of random encoders satisfying \eqref{eq:Condition1OfEncoders} and \eqref{eq:Condition2OfEncoders}. If it holds that
\begin{IEEEeqnarray}{l}
\Pr\{\Phi_{nN}^{(i)}(\seq{v}_i) = \seq{x}_i\} = \prod_{j=0}^{n-1} \Pr\{\Phi_{N}^{(i)}(v_{i,jN+1\cdots (j+1)N}) \IEEEnonumber \\
\qquad = x_{i, jN+1\cdots (j+1)N}\}, \qquad \forall i \in \mathcal{I}_K
\end{IEEEeqnarray}
and
\begin{IEEEeqnarray}{l}
H(V_A|V_{A^c}) < \frac{1}{N} I(X_A^N; Y^N|X_{A^c}^N, V_{A^c}^N) \IEEEnonumber \\
\qquad - \frac{1}{N} \plimsup_{n \to \infty} \rho(\Phi_{nN}^{(A)})(V_A^{nN}, X_A^{nN}) \label{eq:AchievableCondition2}
\end{IEEEeqnarray}
for some positive integer $N$ and all nonempty set $A \subseteq \mathcal{I}_K$, then there exists an $(nN, \epsilon_{nN})$ code based on $(\Phi_{nN}^{(i)})_{i \in \mathcal{I}_K}$ such that $V_{\mathcal{I}_K}$ is transmissible over $W$, where the probability distributions
\begin{IEEEeqnarray}{l}
P_{V_{\mathcal{I}_K}^NX_{\mathcal{I}_K}^NY^N}(\seq{v}_{\mathcal{I}_K}, \seq{x}_{\mathcal{I}_K}, \seq{y}) = P_{V_{\mathcal{I}_K}^N}(\seq{v}_{\mathcal{I}_K}) \IEEEnonumber \\
\qquad P_{Y^N|X_{\mathcal{I}_K}^N}(\seq{y}|\seq{x}_{\mathcal{I}_K}) \prod_{i \in \mathcal{I}_K} P_{X_i^N|V_i^N}(\seq{x}_i|\seq{v}_i) \\
P_{X_i^N|V_i^N}(\seq{x}_i | \seq{v}_i) = \Pr\{\Phi_N^{(i)}(\seq{v}_i) = \seq{x}_i\}
\end{IEEEeqnarray}
and
\begin{IEEEeqnarray}{l}
P_{X_i^{nN}|V_i^{nN}}(\seq{x}_i | \seq{v}_i) = \IEEEnonumber \\
\qquad \prod_{j=0}^{n-1} P_{X_i^N|V_i^N}(x_{i, jN+1\cdots (j+1)N} | v_{i, jN+1\cdots (j+1)N}) \label{eq:ConditionalDistributionX|V2}.
\end{IEEEeqnarray}
\end{corollary}

\begin{remark}\label{re:ArtOfJSCC}
If the encoders $(\Phi_n^{(i)})_{i \in \mathcal{I}_K}$ satisfy \eqref{eq:RandomCodeForMAC}, then the condition \eqref{eq:ConditionOfGeneralizedDirectPartForMAC} in Theorem \ref{th:GeneralizedDirectPartForMAC} reduces to the following one:
\begin{IEEEeqnarray*}{Cl}
&\limsup_{n \to \infty} \Pr\biggl\{\bigcup_{A \subseteq \mathcal{I}_K, A \ne \emptyset} \biggl[ i(X_A^n; Y_A^n|X_{A^c}^n, V_{A^c}^n) \\
&\le h(V_A^n | V_{A^c}^n) + \gamma_n \biggr] \biggr\}\\
\eqvar{(a)} &\limsup_{n \to \infty} \Pr\biggl\{ (V_{\mathcal{I}_K}^n, X_{\mathcal{I}_K}^n, Y^n) \in \bigcup_{\scriptstyle A \subseteq \mathcal{I}_K, \atop \scriptstyle A \ne \emptyset} \biggl[T_{n,\gamma_n}^A(\Phi_n)\biggr]^c \biggr\} \\
\eqvar{(b)} &\limsup_{n \to \infty} \Pr\left\{ (V_{\mathcal{I}_K}^n, X_{\mathcal{I}_K}^n, Y^n) \not \in T_{n,\gamma_n}(\Phi_n) \right\} \\
\le &\epsilon
\end{IEEEeqnarray*}
where (a) follows from the definition \eqref{eq:GoodSet2ForDecoding} and (b) from the definition \eqref{eq:GoodSetForDecoding}. Because the construction $(\mathcal{F}_n^{(i)})_{i \in \mathcal{I}_k}$ in Remark \ref{re:1stForFeinsteinLemmaForMAC} ensures the existence of encoders satisfying \eqref{eq:RandomCodeForMAC}, the direct theorem in \cite[Theorem 1]{JSCC:Iwata200511} clearly follows. Analogously, when the sources and channels are stationary and memoryless, if the encoders $(\Phi_{nN}^{(i)})_{i \in \mathcal{I}_K}$ satisfy \eqref{eq:RandomCodeForMAC}, then we have
$$
\plimsup_{n \to \infty} \rho(\Phi_{nN}^{(A)})(V_A^{nN}, X_A^{nN}) = 0.
$$
Hence by considering all conditional probability distributions
$$
\{P_{X_i^N|V_i^N}(\cdot|\cdot)\}_{i \in \mathcal{I_K}}, \qquad N = 1, 2, \cdots
$$
and constructing the encoders $(\mathcal{F}_{nN}^{(i)})_{i \in \mathcal{I}_k}$ according to the conditional probability distribution \eqref{eq:ConditionalDistributionX|V2}, the direct part of Theorem 2 in \cite{JSCC:Cover198011} follows explicitly from Corollary \ref{co:DirectPartForMemorylessMAC}. The above facts tell us that the condition \eqref{eq:RandomCodeForMAC} stand out to be an important characteristic of good encoders. Furthermore, based on the observation at the end of Remark \ref{re:1stForFeinsteinLemmaForMAC}, we find that \emph{one feasible approach to construct lossless JSCC is as simple as finding some mechanism for generating a set of pairwise independent random codewords subject to a good conditional probability distribution $P_{X_i^n|V_i^n}$ of channel input given source output}. This observation is interesting, since in the classical proofs of many theorems in information theory, we usually generate a set of independent random codewords. However, as we have shown in this paper, only pairwise independence is needed in the proof of lossless JSCC problem. In case of channel coding, we have known that the pairwise independence is sufficient (see \cite[Section 6.2]{JSCC:Gallager196800} and \cite{JSCC:Delsarte198207}). Now, such a viewpoint is extended to the case of JSCC.
\end{remark}

In light of the above ideas, we define a tuple $(\Phi_n^{(i)})_{i \in \mathcal{I}_K}$ of encoders to be \emph{good} if it satisfies \eqref{eq:Condition1OfEncoders}, \eqref{eq:Condition2OfEncoders} and
\begin{equation}\label{eq:ConditionOfGoodEncoders}
\sup_{(\seq{v}_A, \seq{x}_A) \in \prod_{i \in A} \mathcal{V}_i^n \times \prod_{i \in A} \mathcal{X}_i^n} \rho(\Phi_n^{(A)})(\seq{v}_A, \seq{x}_A) = 0
\end{equation}
for all nonempty set $A \subseteq \mathcal{I}_K$. Note that it can be easily shown that the inequality
$$
\sup_{(\seq{v}_A, \seq{x}_A) \in \prod_{i \in A} \mathcal{V}_i^n \times \prod_{i \in A} \mathcal{X}_i^n} \rho(\Phi_n^{(A)})(\seq{v}_A, \seq{x}_A) \ge 0
$$
always holds for any $\{\Phi_n^{(i)}\}_{i \in \mathcal{I}_K}$, and in fact, condition \eqref{eq:ConditionOfGoodEncoders} is equivalent to \eqref{eq:RandomCodeForMAC}. In practice, condition \eqref{eq:ConditionOfGoodEncoders} may be too strict, so we define a tuple $(\Phi_n^{(i)})_{i \in \mathcal{I}_K}$ of encoders to be \emph{$\delta$-asymptotically good} ($\delta \ge 0$) if it satisfies \eqref{eq:Condition1OfEncoders}, \eqref{eq:Condition2OfEncoders}, and
\begin{equation}\label{eq:ConditionOfAsymptoticGoodEncoders}
\limsup_{n \to \infty} \sup_{(\seq{v}_A, \seq{x}_A) \in \prod_{i \in A} \mathcal{V}_i^n \times \prod_{i \in A} \mathcal{X}_i^n} \rho(\Phi_n^{(A)})(\seq{v}_A, \seq{x}_A) \le \delta
\end{equation}
for all nonempty set $A \subseteq \mathcal{I}_K$. We also define a tuple of encoders to be \emph{$\delta$-asymptotically good with respect to $\bm{V}_{\mathcal{I}_K}$} if it satisfies \eqref{eq:Condition1OfEncoders}, \eqref{eq:Condition2OfEncoders}, and
\begin{equation}\label{eq:ConditionOfSpecialAsymptoticGoodEncoders}
\plimsup_{n \to \infty} \rho(\Phi_n^{(A)})(V_A^n, X_A^n) \le \delta
\end{equation}
for all nonempty set $A \subseteq \mathcal{I}_K$, where the probability distribution is define by \eqref{eq:DistributionOfVXY}. When $\delta = 0$, we say the encoders are asymptotically good or asymptotically good with respect to some given sources. The relations among the above three definitions are established in the following proposition. The proof is easy and hence omitted.

\begin{proposition}
Good encoders are asymptotically good, and $\delta$-asymptotically good encoders are $\delta$-asymptotically good with respect to any given sources.
\end{proposition}

\subsection{Lossless JSCC Schemes Based on Linear Codes and Random Interleavers}\label{subsec:IIIc}

Now, let us consider the problem of how to construct good encoders. We notice that the random affine function $\hat{F}$ constructed in Proposition \ref{pr:IndependenceLC1} does exhibit some properties of pairwise independence, so it is natural to consider some way of constructing good encoders based on $\hat{F}$. We obtain the following lemma.

\begin{lemma}\label{le:IndependenceLC2}
For a given random linear code $F_n: \mathcal{V}^n \to \mathcal{U}^l$ and a given (quantization) function $q_n: \mathcal{V}^n \times \mathcal{U}^l \to \mathcal{X}^m$, we have
\begin{IEEEeqnarray}{Cl}
&\Pr\{\Phi_{F_n, q_n}(\seq{v}) = \seq{x}\} = P_{X^m|V^n}(\seq{x}|\seq{v}) \\
&\Pr\{\Phi_{F_n, q_n}(\hat{\seq{v}}) = \hat{\seq{x}} | \Phi_{F_n, q_n}(\seq{v}) = \seq{x}\} \IEEEnonumber \\
= &\beta(F_n, q_n)(\seq{v}, \hat{\seq{v}}, \seq{x}, \hat{\seq{x}}) P_{X^m|V^n}(\hat{\seq{x}}|\hat{\seq{v}}) \label{eq:Independence1ofLC2} \\
\le &\beta'(F_n, q_n)(\hat{\seq{v}}, \hat{\seq{x}}) P_{X^m|V^n}(\hat{\seq{x}}|\hat{\seq{v}}) \label{eq:Independence2ofLC2}
\end{IEEEeqnarray}
for any unequal $\seq{v}, \hat{\seq{v}} \in \mathcal{V}^n$ and any $\seq{x}, \hat{\seq{x}} \in \mathcal{X}^m$, where
\begin{IEEEeqnarray}{rCl}
\Phi_{F_n, q_n}(\seq{v}) &\eqdef &q_n(\seq{v}, \hat{F}_n(\seq{v})) \label{eq:DefinitionOfGoodEncoderBasedOnLC} \\
P_{X^m|V^n}(\seq{x}|\seq{v}) &\eqdef &\frac{|q_n^{-1}(\seq{v}, \seq{x})|}{|\mathcal{U}|^l} \label{eq:SimulationOfConditionalProbability} \\
q_n^{-1}(\seq{v}, \seq{x}) &\eqdef &\{\seq{u} \in \mathcal{U}^l | q_n(\seq{v}, \seq{u}) = \seq{x}\}
\end{IEEEeqnarray}
\begin{IEEEeqnarray}{l}
\beta(F_n, q_n)(\seq{v}, \hat{\seq{v}}, \seq{x}, \hat{\seq{x}}) \eqdef \IEEEnonumber \\
\quad \sum_{\scriptstyle \seq{u} \in q_n^{-1}(\seq{v}, \seq{x}), \atop \scriptstyle \hat{\seq{u}} \in q_n^{-1}(\hat{\seq{v}}, \hat{\seq{x}})} \frac{\alpha(F_n)(P_{\hat{\seq{v}}-\seq{v}}, P_{\hat{\seq{u}}-\seq{u}})}{|q_n^{-1}(\seq{v}, \seq{x})||q_n^{-1}(\hat{\seq{v}}, \hat{\seq{x}})|} \label{eq:DefinitionOfBeta} \\
\beta'(F_n, q_n)(\hat{\seq{v}}, \hat{\seq{x}}) \eqdef \IEEEnonumber \\
\quad \max_{\scriptstyle P \in \mathcal{P}_n(\mathcal{V}) \backslash \{P_{0^n}\}, \atop \scriptstyle \seq{u} \in \mathcal{U}^l} \sum_{\hat{\seq{u}} \in q_n^{-1}(\hat{\seq{v}}, \hat{\seq{x}})}  \frac{\alpha(F_n)(P, P_{\hat{\seq{u}}-\seq{u}})}{|q_n^{-1}(\hat{\seq{v}}, \hat{\seq{x}})|}. \label{eq:DefinitionOfBeta'}
\end{IEEEeqnarray}
\end{lemma}

The proof of Lemma \ref{le:IndependenceLC2} is presented in Section \ref{subsec:ProofOfSectionIII}.

\begin{remark}
Setting $m = n$ and comparing \eqref{eq:Independence2ofLC2} with the condition \eqref{eq:Condition2OfEncoders} or \eqref{eq:Condition3OfEncoders}, we immediately have
\begin{equation}\label{eq:RelationBetweenRho&Beta}
\rho(\Phi_{F_n, q_n}) = \frac{1}{n} \ln \beta'(F_n, q_n).
\end{equation}
Recall that in Proposition \ref{pr:GoodLinearCodes}, we proved that the random linear code $F_{q, n, l}^{\mathrm{RLC}}$ satisfies
$$
\alpha_{F_{q, n, l}^{\mathrm{RLC}}}(P, Q) = 1
$$
for all $P \in \mathcal{P}_n(\mathcal{V}) \backslash \{P_{0^n}\}$ and $Q \in \mathcal{P}_l(\mathcal{U})$, where $\mathcal{V} = \mathcal{U} = \mathbb{F}_q$. Hence, we have $\beta'(F_{q, n, l}^{\mathrm{RLC}}, q_n) \equiv 1$ or $\rho(\Phi_{F_{q, n, l}^{\mathrm{RLC}}, q_n}) \equiv 0$, which means that $\Phi_{F_{q, n, l}^{\mathrm{RLC}}, q_n}$ is a good encoder, that is, it can perform as good as the encoders $(\mathcal{F}_n^{(i)})_{i \in \mathcal{I}_k}$. Furthermore, according to Remark \ref{re:1stForFeinsteinLemmaForMAC}, we may construct a $K$-tuple of good encoders satisfying \eqref{eq:Condition3OfEncoders} based on a $K$-tuple of independent random linear codes, or more specifically, we may only use one deterministic linear code $K$ times, as long as it is good enough and its alphabet size is greater than or equal to the maximum size of the alphabets of all sources.
\end{remark}

According to the relation between $\alpha(F_n)$ and $\rho(\Phi_{F_n, q_n})$ established by \eqref{eq:DefinitionOfBeta'} and \eqref{eq:RelationBetweenRho&Beta}, and following the way of defining good encoders, we define a random linear code $F_n: \mathcal{V}^n \to \mathcal{U}^l$ to be \emph{good} if it satisfies
\begin{equation}\label{eq:ConditionOfGoodLinearCodes}
\max_{\scriptstyle P \in \mathcal{P}_n(\mathcal{V}) \backslash \{P_{0^n}\}, \atop \scriptstyle Q \in \mathcal{P}_l(\mathcal{U})} \alpha(F_n)(P, Q) = 1.
\end{equation}
It can be shown easily by definition \eqref{eq:DefinitionOfAlpha} and identity \eqref{eq:Identity2OfLC} that
$$
\max_{\scriptstyle P \in \mathcal{P}_n(\mathcal{V}) \backslash \{P_{0^n}\}, \atop \scriptstyle Q \in \mathcal{P}_l(\mathcal{U})} \alpha(F_n)(P, Q) \ge 1.
$$
Or else, we may conclude that
\begin{IEEEeqnarray*}{Cl}
&\sum_{\scriptstyle P \in \mathcal{P}_n(\mathcal{V}), \atop \scriptstyle Q \in \mathcal{P}_l(\mathcal{U})} E[S_{\mathcal{V}\mathcal{U}}(F_n)(P, Q)] \\
\eqvar{(a)} &\frac{1}{|\mathcal{V}|^n} + \sum_{\scriptstyle P \in \mathcal{P}_n(\mathcal{V}) \backslash \{P_{0^n}\}, \atop \scriptstyle Q \in \mathcal{P}_l(\mathcal{U})} E[S_{\mathcal{V}\mathcal{U}}(F_n)(P, Q)] \\
\ltvar{(b)} &\frac{1}{|\mathcal{V}|^n} + \sum_{\scriptstyle P \in \mathcal{P}_n(\mathcal{V}) \backslash \{P_{0^n}\}, \atop \scriptstyle Q \in \mathcal{P}_l(\mathcal{U})} S_{\mathcal{V}\mathcal{U}}(\mathcal{V}^n \times \mathcal{U}^l)(P, Q) \\
= &\frac{1}{|\mathcal{V}|^n} + \sum_{\scriptstyle P \in \mathcal{P}_n(\mathcal{V}) \backslash \{P_{0^n}\}} S_{\mathcal{V}}(\mathcal{V}^n)(P) \\
= &1
\end{IEEEeqnarray*}
which contradicts the obvious fact that $E[S_{\mathcal{V}\mathcal{U}}(F_n)(P, Q)]$ is a well-defined joint spectrum. Here, (a) follows from \eqref{eq:DefinitionOfAlpha} and \eqref{eq:Identity2OfLC}, and (b) from the assumption that
$$
\max_{\scriptstyle P \in \mathcal{P}_n(\mathcal{V}) \backslash \{P_{0^n}\}, \atop \scriptstyle Q \in \mathcal{P}_l(\mathcal{U})} \alpha(F_n)(P, Q) < 1.
$$
Further, if condition \eqref{eq:ConditionOfGoodLinearCodes} holds, then we have
\begin{equation}\label{eq:Condition2OfGoodLinearCodes}
\alpha(F_n)(P, Q) = 1
\end{equation}
for all $P \in \mathcal{P}_n(\mathcal{V}) \backslash \{P_{0^n}\}$ and $Q \in \mathcal{P}_l(\mathcal{U})$. This is because if there exists at least one pair $(P, Q)$ ($P \ne P_{0^n}$) such that
$$
\alpha(F_n)(P, Q) < 1
$$
or equivalently
$$
E[S_{\mathcal{V}\mathcal{U}}(F_n)(P, Q)] < S_{\mathcal{V}\mathcal{U}}(\mathcal{V}^n \times \mathcal{U}^l)(P, Q)
$$
then by analogous arguments above, we have
$$
\sum_{\scriptstyle P \in \mathcal{P}_n(\mathcal{X}), \atop \scriptstyle Q \in \mathcal{P}_m(\mathcal{Y})} E[S_{\mathcal{X}\mathcal{Y}}(F_n)(P, Q)] < 1
$$
which again contradicts the fact that $E[S_{\mathcal{X}\mathcal{Y}}(F_n)(P, Q)]$ is a well-defined joint spectrum.

We further define a random linear code $F_n: \mathcal{V}^n \to \mathcal{U}^{l_n}$ to be \emph{$\delta$-asymptotically good} if it satisfies
\begin{equation}\label{eq:ConditionOfAsymptoticGoodLinearCodes}
\limsup_{n \to \infty} \max_{\scriptstyle P \in \mathcal{P}_n(\mathcal{V}) \backslash \{P_{0^n}\}, \atop \scriptstyle Q \in \mathcal{P}_{l_n}(\mathcal{U})} \frac{1}{n} \ln \alpha(F_n)(P, Q) \le \delta.
\end{equation}
When $\delta = 0$, we call it an asymptotically good linear code. A good linear code is evidently asymptotically good.

The relation between good linear codes and good encoders are established by the following proposition. The proof is easy and hence omitted.

\begin{proposition}
If the sequence $\{F_n\}_{n=1}^\infty$ of linear codes $F_n$ is good, then the sequence $\{\Phi_{F_n, q_n}\}_{n=1}^\infty$ of encoders $\Phi_{F_n, q_n}$ defined by \eqref{eq:DefinitionOfGoodEncoderBasedOnLC} is good. If the sequence $\{F_n\}_{n=1}^\infty$ of linear codes is $\delta$-asymptotically good, then the sequence $\{\Phi_{F_n, q_n}\}_{n=1}^\infty$ of encoders is $\delta$-asymptotically good.
\end{proposition}

\begin{figure*}[htbp]
\centering
\includegraphics{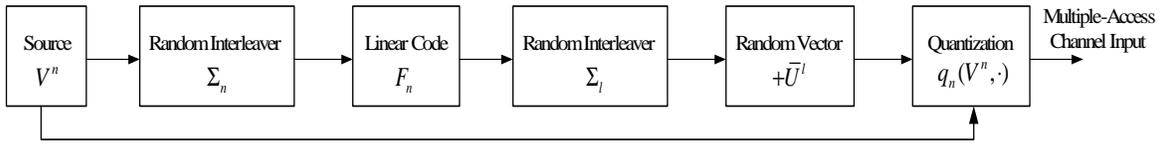}
\caption{The proposed lossless joint source-channel encoding scheme based on linear codes for multiple-access channels}
\label{fig:Scheme1}
\end{figure*}

\begin{figure*}[htbp]
\centering
\subfigure[Channel Coding]{\includegraphics{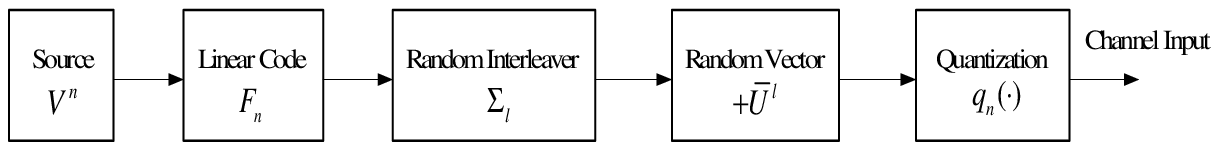}}
\subfigure[Lossless Source Coding]{\includegraphics{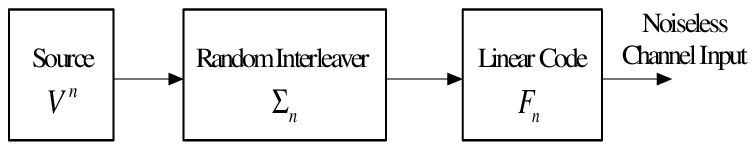}}
\caption{The channel encoding scheme and the lossless source encoding scheme based on linear codes for (multiple-access) channels}
\label{fig:Scheme2}
\end{figure*}

Therefore, Lemma \ref{le:IndependenceLC2} provides a feasible scheme for constructing good encoders based on good or asymptotically good linear codes, which is depicted by Fig. \ref{fig:Scheme1}. Since all encoders work in the same way in the case of multiple terminals, only one encoder is drawn in Fig. \ref{fig:Scheme1} for clarity and simplicity. For comparison, the linear-codes-based channel coding scheme (\cite{JSCC:Gallager196800, JSCC:Bennatan200403}) as well as the linear codes based lossless source coding scheme (\cite{JSCC:Muramatsu200510, JSCC:Yang200503}) are shown in Fig. \ref{fig:Scheme2}. Note that our scheme is in fact the combination of the two schemes except that the quantization $q_n$ is now modified to be directly correlated with the source output $V^n$, which is \emph{the most important characteristic of our scheme}. It demonstrates \emph{the essence of cooperation in a network when sources are correlated or some side information is available}. In practical designs, researchers have realized (e.g., \cite[p. 992]{JSCC:Murugan200408}, \cite[p. 61]{JSCC:Garcia200709}) that preserving the correlation among sources in codewords can facilitate good cooperations among terminals for simultaneous transmission over a MAC, but it is difficult (maybe impossible for some cases) to design a linear code which not only has good distance (or spectrum) properties but also preserves the correlation of sources. In our scheme, such difficulties are perfectly eliminated by adding the quantization module $q_n$, which is directly correlated with the source output, so correlation among sources can be easily preserved in channel inputs without disturbing the design of linear codes. Another important thing to be noted in this scheme is that, a pair of random interleavers is employed in each encoder, an outer interleaver $\Sigma_n$ and an inner interleaver $\Sigma_l$. In general, these interleavers are indispensable to the whole system. As for stationary memoryless sources, supposing a framework with $K$ encoders (thus $K$ pairs of inner and outer interleavers), only one out of the $K$ outer interleavers (could be anyone) can be omitted. Analogously, only one out of the $K$ inner interleavers can be omitted for stationary memoryless MACs. The omitted inner and outer interleavers are not required to belong to the same pair.

According to the proposed scheme illustrated in Fig. \ref{fig:Scheme1}, three main problems needs to be solved in the design of lossless JSCC:

1) \emph{A problem of optimization}. How do we find good (or at least transmissible) conditional probability distributions $(P_{X_i^n|V_i^n})_{i \in \mathcal{I}_K}$ for given correlated sources and MACs?

2) \emph{A problem of coding}%
\footnote{This problem has been solved partly in \cite{JSCC:Yang200808} very recently.}. How do we construct a good linear code \eqref{eq:ConditionOfGoodLinearCodes} or a $\delta$-asymptotically good linear code \eqref{eq:ConditionOfAsymptoticGoodLinearCodes} with arbitrarily small $\delta$, and how do we design the quantization function $q_n$ for a given conditional probability distribution?

3) \emph{A problem of decoding}. How do we decode with low complexities and permissible loss of optimality?

Each of the three problems is difficult and deserves further investigations. We now discuss some related subjects in the following section.

\section{Further Discussions}\label{sec:Discussion}

\subsection{Some Properties of Linear Codes with Good Joint Spectra}\label{subsec:PropertiesOfGoodLinearCodes}

According to the definition \eqref{eq:ConditionOfGoodLinearCodes} or \eqref{eq:Condition2OfGoodLinearCodes} in Section \ref{sec:JSCC} and the identity \eqref{eq:SpectrumOfRandomBin} in Proposition \ref{pr:RandomBin}, a good random linear code $F: \mathcal{X}^n \to \mathcal{Y}^m$ should satisfy
$$
E[S_{\mathcal{X}\mathcal{Y}}(F)(P, Q)] = E[S_{\mathcal{X}\mathcal{Y}}(F_{\mathcal{Y}^m|\mathcal{X}^n}^{\mathrm{RB}})(P, Q)]
$$
for all $P \in \mathcal{P}_n(\mathcal{X}) \backslash \{P_{0^n}\}$ and $Q \in \mathcal{P}_m(\mathcal{Y})$, which tells us that \emph{a good (random) linear code has the same joint spectrum as the random binning function} (when $P \ne P_{0^n}$).

About this observation, we would like to give some comments here. Note that the definition \eqref{eq:ConditionOfGoodLinearCodes} may be rewritten as
$$
\max_{\scriptstyle P \in \mathcal{P}_n(\mathcal{X}) \backslash \{P_{0^n}\}, \atop \scriptstyle Q \in \mathcal{P}_m(\mathcal{Y})} E \biggl[ \frac{S_{\mathcal{X}\mathcal{Y}}(F_n)(P, Q)}{S_{\mathcal{X}\mathcal{Y}}(\mathcal{X}^n \times \mathcal{Y}^m)(P, Q)} \biggr] = 1.
$$
However, it is not equivalent to the following equality
$$
E \biggl[ \max_{\scriptstyle P \in \mathcal{P}_n(\mathcal{X}) \backslash \{P_{0^n}\}, \atop \scriptstyle Q \in \mathcal{P}_m(\mathcal{Y})} \frac{S_{\mathcal{X}\mathcal{Y}}(F_n)(P, Q)}{S_{\mathcal{X}\mathcal{Y}}(\mathcal{X}^n \times \mathcal{Y}^m)(P, Q)} \biggr] = 1.
$$
So we can not claim that there exists a single good linear code $f$ in the sense of \eqref{eq:ConditionOfGoodLinearCodes} even though there exists a good random linear code $F$ (e.g., $F_{q,n,m}^{\mathrm{RLC}}$). To further explore the relation between a good random linear code and its sample codes, we prove the following proposition.

\begin{proposition}\label{pr:RelationBetweenRLC&SLC}
Let $F_n: \mathcal{X}^n \to \mathcal{Y}^{m_n}$ be a random linear code whose rate satisfies $\sup_{n \ge 1} R(F_n) < \infty$. If $F_n$ is good, there exists at least one sample code $f_n$ such that
\begin{IEEEeqnarray}{Cl}
&\max_{\scriptstyle P \in \mathcal{P}_n(\mathcal{X}) \backslash \{P_{0^n}\}, \atop \scriptstyle Q \in \mathcal{P}_{m_n}(\mathcal{Y})} \frac{S_{\mathcal{X}\mathcal{Y}}(f_n)(P, Q)}{S_{\mathcal{X}\mathcal{Y}}(\mathcal{X}^n \times \mathcal{Y}^{m_n})(P, Q)} \IEEEnonumber \\
< &(n+1)^{c_1} (m_n+1)^{c_2} \label{eq:RelationBetwwenRLC&SLC}
\end{IEEEeqnarray}
where $c_1$ and $c_2$ are arbitrary constants satisfying $c_1 > |\mathcal{X}|$ and $c_2 > |\mathcal{Y}|$, which implies that $f_n$ is asymptotically good. Analogously, if $F_n$ is $\delta$-asymptotically good, then there exists at least one sample code $f_n$ which is $\delta$-asymptotically good.
\end{proposition}

The proof of Proposition \ref{pr:RelationBetweenRLC&SLC} is presented in Section \ref{subsec:ProofOfSectionIV}.

Next, to have deeper understanding of good linear codes, we establish the following theorem on the ``distance'' property of good linear codes.

\begin{theorem}\label{th:DistanceProperty}
For a given sequence $\bm{F} = \{F_n\}_{n=1}^\infty$ of $\delta$-asymptotically good linear codes $F_n: \mathcal{X}^n \to \mathcal{Y}^{m_n}$, we have
\begin{equation}\label{eq:DistanceProperty}
\lim_{n \to \infty} E[|B(F_n, h_{\mathcal{X}}, h_{\mathcal{Y}})|] = 0
\end{equation}
for any real numbers $h_{\mathcal{X}}$ and $h_{\mathcal{Y}}$ satisfying
\begin{equation}\label{eq:DistanceInequality}
(h_{\mathcal{X}} + \delta)\overline{R}(\bm{F}) + h_{\mathcal{Y}} < \ln|\mathcal{Y}|
\end{equation}
where
\begin{IEEEeqnarray}{l}
B(F_n, h_{\mathcal{X}}, h_{\mathcal{Y}}) \eqdef \{(\seq{x}, \seq{y}) \in \mathrm{rl}(F_n) | \IEEEnonumber \\
\qquad \seq{x} \ne 0^n, H(P_{\seq{x}}) \le h_{\mathcal{X}}, H(P_{\seq{y}}) \le h_{\mathcal{Y}}\} \label{eq:DefinitionOfB}
\end{IEEEeqnarray}
and
\begin{equation}\label{eq:RateOfLinearCode}
\overline{R}(\bm{F}) \eqdef \limsup_{n \to \infty} R(F_n)
\end{equation}
is called the sup-rate of the code sequence $\bm{F}$ (we assume that $\overline{R}(\bm{F}) < \infty$). Furthermore, we have
\begin{IEEEeqnarray}{Cl}
&\pliminf_{n \to \infty} [(\ln |\mathcal{X}| + \delta) \overline{R}(\bm{F}) + \min_{\seq{x} \ne 0^n} H(P_{F_n(\seq{x})})] \IEEEnonumber \\
\ge &\pliminf_{n \to \infty} \min_{\seq{x} \ne 0^n} [(H(P_{\seq{x}}) + \delta) \overline{R}(\bm{F}) + H(P_{F_n(\seq{x})})] \IEEEnonumber \\
\ge &\ln|\mathcal{Y}|. \label{eq:DistanceProperty2}
\end{IEEEeqnarray}
\end{theorem}

The proof of Theorem \ref{th:DistanceProperty} is presented in Section \ref{subsec:ProofOfSectionIV}.

\begin{remark}
The quantity $H(P_{\seq{x}})$ in Theorem \ref{th:DistanceProperty} is just the well-known Shannon entropy defined by
$$
H(P) \eqdef \sum_{a \in \mathcal{X}} P(a) \ln \frac{1}{P(a)}
$$
and here it is used as a substitute for Hamming weight in the binary case. Unlike Hamming weight, entropy is not a norm, but it does possess the essential properties that we need to describe a code. From the second inequality of \eqref{eq:DistanceProperty2}, we find that, for asymptotically good linear codes ($\delta = 0$), \emph{all the input sequences with low entropy except the all-zero sequence are mapped into output sequences with high entropy}. Next, let $\mathcal{X} = \mathcal{Y} = \mathbb{F}_q$ and $F_n$ be the random linear code $F_{q, n, m_n}^{\mathrm{RLC}}$ defined in Proposition \ref{pr:GoodLinearCodes}, then it follows from \eqref{eq:DistanceProperty2} that
$$
\pliminf_{n \to \infty} \min_{\seq{x} \ne 0^n} H(P_{F_{q, n, m_n}^{\mathrm{RLC}}\compact(\seq{x})}) \ge (1 - \overline{R}(\{F_{q, n, m_n}^{\mathrm{RLC}}\}_{n=1}^\infty)) \ln q.
$$
Let $Y^{m_n}$ be the codeword satisfying
$$
P_{Y^{m_n}}(0) = \max_{\seq{x} \ne 0^n} P_{F_{q, n, m_n}^{\mathrm{RLC}}(\seq{x})}(0)
$$
and define
$$
\Delta_n \eqdef 1 - P_{Y^{m_n}}(0).
$$
Clearly, it is just the normalized distance of the code $F_{q, n, m_n}^{\mathrm{RLC}}$. Then we have
\begin{IEEEeqnarray}{Cl}
&\pliminf_{n \to \infty} [h(\Delta_n) + \Delta_n \ln (q-1)] \IEEEnonumber \\
= &\pliminf_{n \to \infty} [h(P_{Y^{m_n}}(0)) + (1-P_{Y^{m_n}}(0)) \ln (q-1)] \IEEEnonumber \\
\gevar{(a)} &\pliminf_{n \to \infty} H(P_{Y^{m_n}}) \IEEEnonumber \\
\ge &\pliminf_{n \to \infty} \min_{\seq{x} \ne 0^n} H(P_{F_{q, n, m_n}^{\mathrm{RLC}}(\seq{x})}) \IEEEnonumber \\
\ge &(1 - \overline{R}(\{F_{q, n, m_n}^{\mathrm{RLC}}\}_{n=1}^\infty)) \ln q \label{eq:GilbertVarshamovBound}
\end{IEEEeqnarray}
where $h(x) = - x \ln x - (1-x) \ln (1-x)$ and (a) follows from the simple inequality
$$
H(P) \le h(P(0)) + (1 - P(0)) \ln (q-1).
$$
The inequality \eqref{eq:GilbertVarshamovBound} is just the well-known asymptotic Gilbert-Varshamov bound (see \cite{JSCC:Gilbert195210} and \cite[Section 2.10.6]{JSCC:Huffman200300}).
\end{remark}

By Theorem \ref{th:DistanceProperty}, we further prove the following interesting results. Their proofs are presented in Section \ref{subsec:ProofOfSectionIV}.

\begin{corollary}\label{co:SystematicCode}
A code $F: \mathcal{X}^n \to \mathcal{X}^m$ ($m \ge n$) is said to be \emph{systematic} if all symbols of the input sequence are directly included in the output sequence. Then, for any given sequence $\bm{F} = \{F_n\}_{n=1}^\infty$ of asymptotically good linear codes $F_n: \mathcal{X}^n \to \mathcal{X}^{m_n}$, we have
\begin{equation}
\overline{R}(\bm{F}) \le \frac{1}{|\mathcal{X}|}
\end{equation}
if these codes are systematic.
\end{corollary}

\begin{corollary}\label{co:SparseMatrix}
Let $\mathcal{X}$ be a finite field, and $A_{n \times m}$ be a random $n \times m$ matrix over $\mathcal{X}$. The average density $D(A_{n \times m})$ of the matrix $A_{n \times m}$ is defined by
\begin{equation}\label{eq:DefinitionOfMatrixDensity}
D(A_{n \times m}) \eqdef \frac{1}{m n} E[\sum_{i=1}^{n} \sum_{j=1}^m 1\{A_{n \times m}(i, j) \ne 0\}]
\end{equation}
where $A_{n \times m}(i, j)$ denotes the element at the $i$-th row and $j$-th column in $A_{n \times m}$. Then we conclude that the sequence of codes defined by $F_n(\seq{x}) = \seq{x} A_{n \times m_n}$ is not asymptotically good if
\begin{equation}\label{eq:MatrixDensityCondition}
\liminf_{n \to \infty} D(A_{n \times m_n}) < 1 - \frac{1}{|\mathcal{X}|}.
\end{equation}
\end{corollary}

As we can see, when $\mathcal{X}$ is a finite filed, Corollary \ref{co:SystematicCode} is virtually an easy consequence of Corollary \ref{co:SparseMatrix}, and the latter establishes the relation between the density of a generator matrix and its joint spectrum. More interestingly, it demonstrates that any sparse generator matrices cannot yield linear codes with good joint spectra. This observation shows an important difference between the traditional design of channel codes and the design of JSCC. In the former case, it is the image of the encoding map (the output codewords) that serves as major concern, while in the latter case, it is the encoding map itself (the generator matrix) rather than its image that determines the performance. So the generator matrix deserves careful examination in the design of JSCC, even for those good linear channel codes in traditional meaning.

However, it is unnecessary to take a pessimistic attitude toward the construction of good linear codes for JSCC. After all, only the generator matrix of a linear code is required to be dense, and we may still construct good linear codes for JSCC based on sparse matrices (e.g., sparse parity-check matrices). In our another paper \cite{JSCC:Yang200808}, we will investigate the problem of constructing linear codes with good joint spectra in detail, and an explicit construction based on sparse matrices will be given.

\subsection{Arbitrary Rate Coding Schemes}\label{subsec:VariableRateCodingSchemes}

In section \ref{sec:JSCC}, we only consider the case when one symbol is transmitted per channel use, but the results established there can be easily extended to more general cases, including correlated sources with different sampling rates and MACs with different transmission rates.

To define correlated sources with different sampling rates, we only need to replace the source $\bm{V}_{\mathcal{I}_K} = \{(V_i^n)_{i \in \mathcal{I}_K}\}_{n=1}^\infty$ by an infinite sequence $\bm{V}^{(\mathcal{I}_K)} = \{(V_n^{(i)})_{i \in \mathcal{I}_K}\}_{n=1}^\infty$ of random vectors and let the alphabet of each $V_n^{(i)}$ be $\mathcal{V}_i^{l_{n,i}}$. Analogously, we may define a MAC with different transmission rates by an infinite sequence $\bm{W} = \{W^n\}_{n=1}^\infty$ of conditional probability distribution $W^n: \prod_{i \in \mathcal{I}_K} \mathcal{X}_i^{m_{n,i}} \to \mathcal{Y}^n$ satisfying
$$
\sum_{\seq{y} \in \mathcal{Y}^n} W^n(y^n|(\seq{x}_i)_{i \in \mathcal{I}_K}) = 1
$$
for all $(\seq{x}_i)_{i \in \mathcal{I}_K} \in \prod_{i \in \mathcal{I}_K} \mathcal{X}_i^{m_{n, i}}$ and each $n = 1, 2, \ldots$. Accordingly, the encoder and the decoder are then modified to be the maps $\phi_n^{(i)}: \mathcal{V}_i^{l_{n,i}} \to \mathcal{X}_i^{m_{n,i}}$ ($i \in \mathcal{I}_K$) and $\psi_n: \mathcal{Y}^n \to \prod_{i \in \mathcal{I}_K} \mathcal{V}_i^{l_{n,i}}$, respectively.

Then, for given correlated general sources $\bm{V}^{(\mathcal{I}_K)}$ and general MAC $\bm{W}$, we define the sources $\bm{V}^{(\mathcal{I}_K)}$ being $(R_{\mathcal{I}_K}, \epsilon)$-transmissible over the channel $\bm{W}$ if there exists an $(n, \epsilon_n)$ code satisfying
$$
\limsup_{n \to \infty} \frac{l_{n,i}}{m_{n,i}} \le R_i, \qquad \forall i \in \mathcal{I}_K
$$
and
$$
\limsup_{n \to \infty} \epsilon_n \le \epsilon
$$
where $R_{\mathcal{I}_K} \eqdef (R_i)_{i \in \mathcal{I}_K}$ denotes a $K$-tuple of nonnegative real numbers.

All the results in Section \ref{sec:JSCC} still hold for the new general sources and channels, but now we will only concentrate on a special case in which $l_{n,i} = l_n$ and $m_{n,i} = n$ for all $i \in \mathcal{I}_K$. In such case, we may define the sources $\bm{V}_{\mathcal{I}_K} = \{(V_i^{l_n})_{i \in \mathcal{I}_K}\}_{n=1}^\infty$ being $(R, \epsilon)$-transmissible over the channel $\bm{W} = \{W^n: \prod_{i \in \mathcal{I}_K} \mathcal{X}_i^n  \to \mathcal{Y}^n\}_{n=1}^\infty$ if there exists an $(n, \epsilon_n)$ code satisfying
$$
\limsup_{n \to \infty} \frac{l_n}{n} \le R, \quad \limsup_{n \to \infty} \epsilon_n \le \epsilon.
$$
In this setting, Lemma \ref{le:FeinsteinLemmaForMAC} and Theorem \ref{th:GeneralizedDirectPartForMAC} in Section \ref{sec:JSCC} can then be restated as follows by Lemma \ref{le:FeinsteinLemmaXForMAC} and Theorem \ref{th:GeneralizedDirectPartXForMAC}, respectively.

\begin{lemma}\label{le:FeinsteinLemmaXForMAC}
Let $\bm{V}_{\mathcal{I}_K}$ be a $K$-tuple of correlated general sources and $\bm{W}$ a MAC with $K$ input terminals. For a given tuple $(\Phi_n^{(i)})_{i \in \mathcal{I}_K}$ of random encoders, if
\begin{equation}\label{eq:Condition1XOfEncoders}
\Pr\{(\Phi_n^{(i)}(\seq{v}_i))_{i \in \mathcal{I}_K} \compact = \seq{x}_{\mathcal{I}_K}\} = \prod_{i \in \mathcal{I}_K} \compact \Pr\{\Phi_n^{(i)}(\seq{v}_i) = \seq{x}_i\}
\end{equation}
for all $\seq{v}_{\mathcal{I}_K} \in \prod_{i \in \mathcal{I}_K} \mathcal{V}_i^{l_n}$ and $\seq{x}_{\mathcal{I}_K} \in \prod_{i \in \mathcal{I}_K} \mathcal{X}_i^n$, and there exits a family of real functions
$$
\bigl\{ \rho(\Phi_n^{(A)}): \prod_{i \in \mathcal{I}_K} \mathcal{V}_i^{l_n} \times \prod_{i \in \mathcal{I}_K} \mathcal{X}_i^n \to \mathbb{R} \bigr\}_{A \subseteq \mathcal{I}_K, A \ne \emptyset}
$$
satisfying
\begin{IEEEeqnarray}{Cl}
&\Pr\{(\Phi_n^{(i)}(\hat{\seq{v}}_i))_{i \in A} = \hat{\seq{x}}_A | (\Phi_n^{(i)}(\seq{v}_i))_{i \in \mathcal{I}_K} = \seq{x}_{\mathcal{I}_K}\} \IEEEnonumber \\
\le &e^{n\rho(\Phi_n^{(A)})(\hat{\seq{v}}_A, \hat{\seq{x}}_A)} \prod_{i \in A} \Pr\{\Phi_n^{(i)}(\hat{\seq{v}}_i) = \hat{\seq{x}}_i\} \label{eq:Condition2XOfEncoders}
\end{IEEEeqnarray}
for all nonempty set $A \subseteq \mathcal{I}_K$, any unequal $\seq{v}_i, \hat{\seq{v}}_i \in \mathcal{V}_i^{l_n}$ ($i \in \mathcal{I}_K$), and any $\seq{x}_i, \hat{\seq{x}}_i \in \mathcal{X}_i^n$ ($i \in \mathcal{I}_K$), then there exists a joint decoder $\Psi_n$ such that the decoding error probability
\begin{equation}
\epsilon_n \le \Pr\left\{ (V_{\mathcal{I}_K}^{l_n}, X_{\mathcal{I}_K}^n, Y^n) \not \in T_{n,\gamma}(\Phi_n) \right\} + (2^K - 1)e^{-n\gamma}
\end{equation}
for all $n = 1, 2, \cdots$, where $\gamma > 0$ and
\begin{IEEEeqnarray}{l}
P_{V_{\mathcal{I}_K}^{l_n}X_{\mathcal{I}_K}^nY^n}(\seq{v}_{\mathcal{I}_K}, \seq{x}_{\mathcal{I}_K}, \seq{y}) = \IEEEnonumber \\
\quad P_{V_{\mathcal{I}_K}^{l_n}}(\seq{v}_{\mathcal{I}_K}) W^n(\seq{y} | \seq{x}_{\mathcal{I}_K}) \prod_{i \in \mathcal{I}_K} P_{X_i^n|V_i^{l_n}}(\seq{x}_i | \seq{v}_i) \label{eq:DistributionXOfVXY} \\
P_{X_i^n|V_i^{l_n}}(\seq{x}_i | \seq{v}_i) = \Pr\{\Phi_n^{(i)}(\seq{v}_i) = \seq{x}_i\} 
\end{IEEEeqnarray}
\begin{IEEEeqnarray}{rCl}
T_{n,\gamma}(\Phi_n) &= &\bigcap_{A \subseteq \mathcal{I}_K, A \ne \emptyset} T_{n,\gamma}^A(\Phi_n) \label{eq:GoodSetXForDecoding} \\
T_{n,\gamma}^A(\Phi_n) &= &\biggl\{ (\seq{v}_{\mathcal{I}_K}, \seq{x}_{\mathcal{I}_K}, \seq{y}) \in \prod_{i \in \mathcal{I}_K} \mathcal{V}_i^{l_n} \times \prod_{i \in \mathcal{I}_K} \mathcal{X}_i^n \times \mathcal{Y}^n \bigg| \IEEEnonumber \\
& &i(\seq{x}_A; \seq{y} | \seq{x}_{A^c}, \seq{v}_{A^c}) > h(\seq{v}_A | \seq{v}_{A^c}) \IEEEnonumber \\
& &+\: \rho(\Phi_n^{(A)})(\seq{v}_A, \seq{x}_A) + \gamma \biggr\} 
\end{IEEEeqnarray}
and
\begin{IEEEeqnarray*}{rCl}
i(\seq{x}_A; \seq{y}|\seq{x}_{A^c}, \seq{v}_{A^c}) &\eqdef &\frac{1}{n} \ln \frac{W^n(\seq{y}|\seq{x}_{\mathcal{I}_K})}{P_{Y^n|X_{A^c}^nV_{A^c}^{l_n}}(\seq{y}|\seq{x}_{A^c}, \seq{v}_{A^c})} \\
h(\seq{v}_A|\seq{v}_{A^c}) &\eqdef &\frac{1}{n} \ln \frac{1}{P_{V_A^{l_n}|V_{A^c}^{l_n}}(\seq{v}_A|\seq{v}_{A^c})}.
\end{IEEEeqnarray*}
\end{lemma}

\begin{theorem}\label{th:GeneralizedDirectPartXForMAC}
Let $\bm{V}_{\mathcal{I}_K}$ be a $K$-tuple of correlated general sources, $\bm{W}$ a general MAC with $K$ input terminals, and $(\Phi_n^{(i)})_{i \in \mathcal{I}_K}$ a given tuple of random encoders satisfying \eqref{eq:Condition1XOfEncoders} and \eqref{eq:Condition2XOfEncoders}. If for a sequence $\{\gamma_n\}_{n=1}^\infty$ satisfying
$$
\gamma_n > 0, \gamma_n \to 0, \mbox{ and } n\gamma_n \to \infty, \qquad \mbox{as } n \to \infty
$$
it holds that
$$
\limsup_{n \to \infty} \frac{l_n}{n} \le R
$$
and
$$
\limsup_{n \to \infty} \Pr\left\{ (V_{\mathcal{I}_K}^{l_n}, X_{\mathcal{I}_K}^n, Y^n) \not \in T_{n,\gamma_n}(\Phi_n) \right\} \le \epsilon
$$
then there exists an $(n, \epsilon_n)$ code based on $(\Phi_n^{(i)})_{i \in \mathcal{I}_K}$ such that $\bm{V}_{\mathcal{I}_K}$ is $(R, \epsilon)$-transmissible over $\bm{W}$, where the probability distribution of $(V_{\mathcal{I}_K}^{l_n}, X_{\mathcal{I}_K}^n, Y^n)$ is defined by \eqref{eq:DistributionXOfVXY}, and $T_{n,\gamma_n}(\Phi_n)$ is defined by \eqref{eq:GoodSetXForDecoding}.
\end{theorem}

A special case of Theorem \ref{th:GeneralizedDirectPartXForMAC}, i.e., the point-to-point case, has been presented in \cite{JSCC:Yang200610A}. For readers familiar with information-spectrum methods, it seems that Theorem \ref{th:GeneralizedDirectPartXForMAC} brings forth no new groundbreaking results compared with Theorem \ref{th:GeneralizedDirectPartForMAC}. However, when considering the concrete design of encoders with multiple rates under our scheme, some interesting results are observed. According to Theorem \ref{th:GeneralizedDirectPartXForMAC}, a feasible approach to lossless JSCC for general rate settings is still to generate a set of pairwise independent random codewords subject to a good conditional probability distribution except that the original conditional probability distribution $P_{X^n|V^n}$ is now replaced by $P_{X^m|V^n}$ ($m$ may not equal $n$). Again by Lemma \ref{le:IndependenceLC2}, an encoder can be defined by \eqref{eq:DefinitionOfGoodEncoderBasedOnLC}. Note that in \eqref{eq:DefinitionOfGoodEncoderBasedOnLC}, if we are given a good linear code $F_n: \mathcal{V}^n \to \mathcal{U}^l$ with very low rate, that is, $R(F_n) \to 0$ or $l / n \gg 1$, then for most conditional probability distributions $P_{X^m|V^n}(\seq{x}|\seq{v})$ with $m$ not too large, the only thing that we need to do is simply to select an appropriate function $q_n$ satisfying \eqref{eq:SimulationOfConditionalProbability}.

This is an interesting conclusion. In traditional designs of channel codes, researchers are usually concerned about how to design a code for each rate required by each specific scenario, but now, a good low-rate code is enough, and all that we need is to design a good quantization function for each rate. We call such a method ``\emph{generalized puncturing}'', but there is a notable difference between our generalized puncturing and the traditional puncturing method. While puncturing is traditionally intended for generating a good code with new higher rates, the goal of generalized puncturing is to generate a good input sequence with multiple rates for a specific channel. For some special cases, e.g., the binary symmetric channel, the above two goals coincide, but for most cases they do not.

\begin{figure*}[htbp]
\centering
\includegraphics{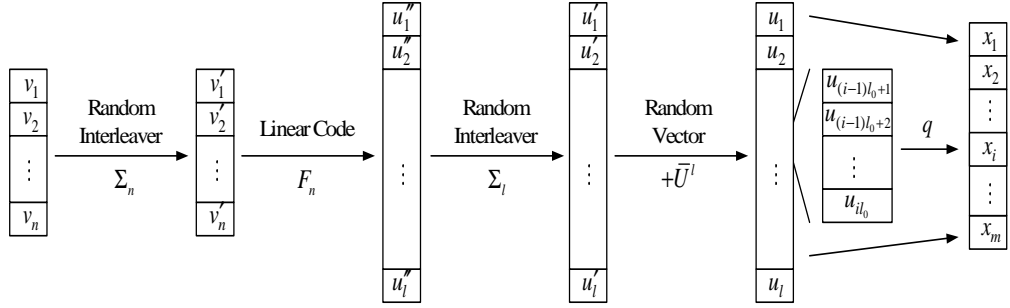}
\caption{The mechanism of generalized puncturing for each encoder when $P_{X^m|V^n}(x^m|v^n) = \prod_{i=1}^m P_X(x)$}
\label{fig:Puncturing}
\end{figure*}

The generalized puncturing works more like the traditional puncturing when the conditional probability distribution satisfies
$$
P_{X^m|V^n}(\seq{x}|\seq{v}) = \prod_{i=1}^m P_X(x_i)
$$
which is the usual case for the point-to-point transmission over a memoryless channel. Let $q: \mathcal{U}^{l_0} \to \mathcal{X}$ be the function satisfying
$$
P_X(a) = \frac{q^{-1}(a)}{|\mathcal{U}|^{l_0}}
$$
for all $a \in \mathcal{X}$. Then we may define the encoder by
$$
\Phi_n(\seq{v}) = q(u_{1 \cdots l_0}) q(u_{l_0+1 \cdots 2l_0}) \cdots q(u_{(m-1)l_0+1 \cdots ml_0})
$$
where $\seq{v} \in \mathcal{V}^n$ and $\seq{u} = \hat{F_n}(\seq{v}) \in \mathcal{U}^l$. Note that $l \ge m l_0$ since $R(F_n)$ is assumed to be very small. This mechanism is visualized in Fig. \ref{fig:Puncturing}.
Furthermore, if $\mathcal{V} = \mathcal{U} = \mathcal{X} = \mathbb{F}_2$ and $P_X(0) = P_X(1) = 0.5$, we may define the function $q$ to be the identity function on $\mathbb{F}_2$, and then our generalized puncturing is almost the same as the traditional puncturing. The only difference is that the generalized puncturing operates on the randomly interleaved coset codes of $F_n$ (i.e., $\hat{F}_n$) instead of $F_n$ itself.

\section{Proofs}\label{sec:Proofs}

\subsection{Proofs of Results in Section \ref{sec:Spectrum}}\label{subsec:ProofOfSectionII}

\begin{proofof}{Proposition \ref{pr:RandomBin}}
By the definition of $F_{\mathcal{Y}^m | \mathcal{X}^n}^{\mathrm{RB}}$, we have
$$
\Pr\{F_{\mathcal{Y}^m | \mathcal{X}^n}^{\mathrm{RB}}(\seq{x}) = \seq{y}\} = |\mathcal{Y}|^{-m}
$$
for any $\seq{x} \in \mathcal{X}^n$ and $\seq{y} \in \mathcal{Y}^m$. Hence, for any $P \in \mathcal{P}_n(\mathcal{X})$ and any $Q \in \mathcal{P}_m(\mathcal{Y})$, we have
\begin{IEEEeqnarray*}{Cl}
&E[S_{\mathcal{X}\mathcal{Y}}(\mathrm{rl}(F_{\mathcal{Y}^m | \mathcal{X}^n}^{\mathrm{RB}}))(P, Q)] \\
= &E\biggl[\frac{1}{|\mathrm{rl}(F_{\mathcal{Y}^m | \mathcal{X}^n}^{\mathrm{RB}})|} \sum_{\scriptstyle \seq{x} \in \mathcal{T}_P^n(\mathcal{X}), \atop \scriptstyle \seq{y} \in \mathcal{T}_Q^m(\mathcal{Y})} 1\{F_{\mathcal{Y}^m | \mathcal{X}^n}^{\mathrm{RB}}(\seq{x}) = \seq{y}\}\biggr] \\
= &\frac{1}{|\mathcal{X}|^n} \sum_{\scriptstyle \seq{x} \in \mathcal{T}_P^n(\mathcal{X}), \atop \scriptstyle \seq{y} \in \mathcal{T}_Q^m(\mathcal{Y})} \Pr\{F_{\mathcal{Y}^m | \mathcal{X}^n}^{\mathrm{RB}}(\seq{x}) = \seq{y}\} \\
= &\frac{|\mathcal{T}_P^n(\mathcal{X})| |\mathcal{T}_Q^m(\mathcal{Y})|}{|\mathcal{X}|^n |\mathcal{Y}|^m}.
\end{IEEEeqnarray*}
This together with Proposition \ref{pr:SpectrumOfSets} concludes \eqref{eq:SpectrumOfRandomBin}, and the identity \eqref{eq:Spectrum2OfRandomBin} can be proved analogously.
\end{proofof}

\begin{proofof}{Proposition \ref{pr:SpectrumPropertyOfFunctions}}
By Proposition \ref{pr:SpectrumInvarianceUnderPermutation}, we have
$$
E[S_{\mathcal{X}\mathcal{Y}}(F)(P, Q)] = E[S_{\mathcal{X}\mathcal{Y}}(\tilde{F})(P, Q)]
$$
for any $P \in \mathcal{P}_n(\mathcal{X})$, $Q \in \mathcal{P}_m(\mathcal{Y})$. Moreover, it is easy to show that
$$
\Pr\{\tilde{F}(\seq{x}_1) = \seq{y}_1\} = \Pr\{\tilde{F}(\seq{x}_2) = \seq{y}_2\}
$$
if $P_{\seq{x}_1} = P_{\seq{x}_2}$ and $P_{\seq{y}_1} = P_{\seq{y}_2}$. Therefore, we have
\begin{IEEEeqnarray*}{Cl}
&E[S_{\mathcal{X}\mathcal{Y}}(F)(P_{\seq{x}}, P_{\seq{y}})] \\
= &E[S_{\mathcal{X}\mathcal{Y}}(\tilde{F})(P_{\seq{x}}, P_{\seq{y}})] \\
= &E[S_{\mathcal{X}\mathcal{Y}}(\mathrm{rl}(\tilde{F}))(P_{\seq{x}}, P_{\seq{y}})] \\
= &E \biggl[ \frac{1}{|\mathrm{rl}(\tilde{F})|} \sum_{\hat{\seq{x}} \in \mathcal{T}_{P_{\seq{x}}}^n(\mathcal{X})} \sum_{\hat{\seq{y}} \in \mathcal{T}_{P_{\seq{y}}}^m(\mathcal{Y})} 1\{\tilde{F}(\hat{\seq{x}}) = \hat{\seq{y}}\} \biggr] \\
= &\frac{1}{|\mathcal{X}|^n} \sum_{\hat{\seq{x}} \in \mathcal{T}_{P_{\seq{x}}}^n(\mathcal{X})} \sum_{\hat{\seq{y}} \in \mathcal{T}_{P_{\seq{y}}}^m(\mathcal{Y})} \Pr\{\tilde{F}(\hat{\seq{x}}) = \hat{\seq{y}}\} \\
= &\frac{1}{|\mathcal{X}|^n} \sum_{\hat{\seq{x}} \in \mathcal{T}_{P_{\seq{x}}}^n(\mathcal{X})} \sum_{\hat{\seq{y}} \in \mathcal{T}_{P_{\seq{y}}}^m(\mathcal{Y})} \Pr\{\tilde{F}(\seq{x}) = \seq{y}\} \\
= &\frac{|\mathcal{T}_{P_{\seq{x}}}^n(\mathcal{X})| |\mathcal{T}_{P_{\seq{y}}}^m(\mathcal{Y})|}{|\mathcal{X}|^n} \Pr\{\tilde{F}(\seq{x}) = \seq{y}\} \\
\eqvar{(a)} &|\mathcal{Y}|^m S_{\mathcal{X}\mathcal{Y}}(\mathcal{X}^n \times \mathcal{Y}^m)(P_{\seq{x}}, P_{\seq{y}}) \Pr\{\tilde{F}(\seq{x}) = \seq{y}\}
\end{IEEEeqnarray*}
where (a) follows from Proposition \ref{pr:SpectrumOfSets}. This together with the definition \eqref{eq:DefinitionOfAlpha} yields \eqref{eq:SpectrumPropertyOfFunctions}.
\end{proofof}

\begin{proofof}{Proposition \ref{pr:GoodLinearCodes}}
The identity \eqref{eq:Identity1OfLC} is an obvious fact in abstract algebra and \eqref{eq:Identity2OfLC} follows easily from \eqref{eq:Identity1OfLC} and Proposition \ref{pr:SpectrumPropertyOfFunctions}.

As for the remaining identities, it suffices to show that the random linear code $F_{q, n, m}^{\mathrm{RLC}}$ satisfies \eqref{eq:Identity1OfGLC}, because \eqref{eq:Identity2OfGLC} is an easy consequence of \eqref{eq:Identity1OfGLC} and Proposition \ref{pr:SpectrumPropertyOfFunctions}.

Note that both $aX$ and $X + Y$ are uniform random variables if $X$ and $Y$ are two independent uniform random variables on $\mathbb{F}_q$ and $a$ is a nonzero constant in $\mathbb{F}_q$, and hence \eqref{eq:Identity1OfGLC} clearly follows for all $\seq{x} \in \mathcal{X}^n \backslash \{0^n\}$, $\seq{y} \in \mathcal{Y}^m$.
\end{proofof}

\begin{proofof}{Proposition \ref{pr:IndependenceLC1}}
The identity \eqref{eq:Identity1OfIndependenceLC1} follows easily from the definition \eqref{eq:RandomizedF2} since $\bar{Y}^m$ is an independent uniform random vector on $\mathcal{Y}^m$. To prove \eqref{eq:Identity2OfIndependenceLC1}, we evaluate
\begin{IEEEeqnarray*}{Cl}
&\Pr\{\hat{F}(\seq{x}) = \seq{y}, \hat{F}(\hat{\seq{x}}) = \hat{\seq{y}}\} \\
\eqvar{(a)} &\Pr\{\tilde{F}(\seq{x}) + \bar{Y}^m = \seq{y}, \tilde{F}(\hat{\seq{x}} - \seq{x}) = \hat{\seq{y}} - \seq{y}\} \\
= &\Pr\{\bar{Y}_1^m = \tilde{F}(\seq{x}), \tilde{F}(\hat{\seq{x}} - \seq{x}) = \hat{\seq{y}} - \seq{y}\} \\
= &\sum_{f} P_{\tilde{F}}(f) E\bigl[ 1\{\bar{Y}_1^m = f(\seq{x}), f(\hat{\seq{x}} - \seq{x}) = \hat{\seq{y}} - \seq{y}\} \bigr] \\
= &\sum_{f} P_{\tilde{F}}(f) \Pr\{\bar{Y}_1^m = f(\seq{x})\} 1\{f(\hat{\seq{x}} - \seq{x}) = \hat{\seq{y}} - \seq{y}\} \\
\eqvar{(b)} &|\mathcal{Y}|^{-m} \Pr\{\tilde{F}(\hat{\seq{x}} - \seq{x}) = \hat{\seq{y}} - \seq{y}\}
\end{IEEEeqnarray*}
where $\bar{Y}_1^m \eqdef \seq{y} - \bar{Y}^m$, (a) follows from the definition \eqref{eq:RandomizedF2}, and (b) follows from the fact that $\bar{Y}_1^m$ is also a uniform random vector on $\mathcal{Y}^m$. Then we have
\begin{IEEEeqnarray*}{Cl}
&\Pr\{\hat{F}(\hat{\seq{x}}) = \hat{\seq{y}} | \hat{F}(\seq{x}) = \seq{y}\} \\
= &\frac{\Pr\{\hat{F}(\seq{x}) = \seq{y}, \hat{F}(\hat{\seq{x}}) = \hat{\seq{y}}\}}{\Pr\{\hat{F}(\seq{x}) = \seq{y}\}} \\
= &\Pr\{\tilde{F}(\hat{\seq{x}} - \seq{x}) = \hat{\seq{y}} - \seq{y}\} \\
\eqvar{(a)} &|\mathcal{Y}|^{-m} \alpha(F)(P_{\hat{\seq{x}} - \seq{x}}, P_{\hat{\seq{y}} - \seq{y}})
\end{IEEEeqnarray*}
where (a) follows from Proposition \ref{pr:SpectrumPropertyOfFunctions}, and this concludes \eqref{eq:Identity2OfIndependenceLC1}.
\end{proofof}

\subsection{Proof of Results in Section \ref{sec:JSCC}}\label{subsec:ProofOfSectionIII}

\begin{proofof}{Lemma \ref{le:FeinsteinLemmaForMAC}}
In order to define a decoder $\Psi_n: \mathcal{Y}^m \to \prod_{i \in \mathcal{I}_K} \mathcal{V}_i^n$, we set
\begin{IEEEeqnarray}{rCl}
T_{n,\gamma}(\Phi_n, \seq{v}_{\mathcal{I}_K}) &= &\biggl\{ (\seq{x}_{\mathcal{I}_K}, \seq{y}) \in \prod_{i \in \mathcal{I}_K} \mathcal{X}_i^n \times \mathcal{Y}^n \bigg| \IEEEnonumber \\
& &(\seq{v}_{\mathcal{I}_K}, \seq{x}_{\mathcal{I}_K}, \seq{y}) \in T_{n,\gamma}(\Phi_n) \biggr\} \label{eq:GoodSet3ForDecoding} \\
T_{n,\gamma}^A(\Phi_n, \seq{v}_{\mathcal{I}_K}) &= &\biggl\{ (\seq{x}_{\mathcal{I}_K}, \seq{y}) \in \prod_{i \in \mathcal{I}_K} \mathcal{X}_i^n \times \mathcal{Y}^n \bigg| \IEEEnonumber \\
& &(\seq{v}_{\mathcal{I}_K}, \seq{x}_{\mathcal{I}_K}, \seq{y}) \in T_{n,\gamma}^A(\Phi_n) \biggr\}. \label{eq:GoodSet4ForDecoding}
\end{IEEEeqnarray}
Clearly
$$
T_{n,\gamma}(\Phi_n, \seq{v}_{\mathcal{I}_K}) = \bigcap_{A \subseteq \mathcal{I}_K, A \ne \emptyset} T_{n,\gamma}^A(\Phi_n, \seq{v}_{\mathcal{I}_K}).
$$

Suppose that a channel output $\seq{y} \in \mathcal{Y}^n$ is received, then we define the decoder by $\seq{v}_{\mathcal{I}_K} = \psi_n(\seq{y})$ if there exits a unique $\seq{v}_{\mathcal{I}_K} \in \prod_{i \in \mathcal{I}_K} \mathcal{V}_i^n$ satisfying $((\varphi_n^{(i)}(\seq{v}_i))_{i \in \mathcal{I}_K}, \seq{y}) \in T_{n,\gamma}(\Phi_n, \seq{v}_{\mathcal{I}_K})$, where $(\varphi_n^{(i)})_{i \in \mathcal{I}_K}$ denotes one realization of the tuple $(\Phi_n^{(i)})_{i \in \mathcal{I}_K}$ of random encoders. If there exists no such $\seq{v}_{\mathcal{I}_K}$ or exist more than one such $\seq{v}_{\mathcal{I}_K}$, we define $\psi_n(\seq{y})$ to be an arbitrary element in $\prod_{i \in \mathcal{I}_K} \mathcal{V}_i^n$. Then, for each sample encoder $(\varphi_n^{(i)})_{i \in \mathcal{I}_K}$ generated by $(\Phi_n^{(i)})_{i \in \mathcal{I}_K}$, there is a well-defined decoder $\psi_n$, and we denote by $\Psi_n = (\Psi_n^{(i)})_{i \in \mathcal{I}_K}$ the whole random ensemble of the decoders with respect to the tuple $(\Phi_n^{(i)})_{i \in \mathcal{I}_K}$ of random encoders. The decoding error probability $\epsilon_n$ with respect to the tuple $((\Phi_n^{(i)})_{i \in \mathcal{I}_K}, \Psi_n)$ is then given by
\begin{equation}\label{eq:AverageErrorFormula}
\epsilon_n = \sum_{\seq{v}_{\mathcal{I}_K} \in \prod_{i \in \mathcal{I}_K} \mathcal{V}_i^n} P_{V_{\mathcal{I}_K}^n}(\seq{v}_{\mathcal{I}_K}) \epsilon_n(\seq{v}_{\mathcal{I}_K})
\end{equation}
where $\epsilon_n(\seq{v}_{\mathcal{I}_K})$ denotes the decoding error probability of $K$ sources' output $\seq{v}_{\mathcal{I}_K} \in \prod_{i \in \mathcal{I}_K} \mathcal{V}_i^n$, and it can be bounded above in the following way:
\begin{IEEEeqnarray}{Cl}
&\epsilon_n(\seq{v}_{\mathcal{I}_K}) \IEEEnonumber \\
\le &\Pr\{((\Phi_n^{(i)}(\seq{v}_i))_{i \in \mathcal{I}_K}, Y^n) \not \in T_{n,\gamma}(\Phi_n, \seq{v}_{\mathcal{I}_K})\} \IEEEnonumber \\
&+\: \Pr\biggl\{\bigcup_{\hat{\seq{v}}_{\mathcal{I}_K}: \hat{\seq{v}}_{\mathcal{I}_K} \ne \seq{v}_{\mathcal{I}_K}} \{((\Phi_n^{(i)}(\hat{\seq{v}}_i))_{i \in \mathcal{I}_K}, Y^n) \IEEEnonumber \\
&\in T_{n,\gamma}(\Phi_n, \hat{\seq{v}}_{\mathcal{I}_K})\} \biggr\} \IEEEnonumber \\
\le &\Pr\{((\Phi_n^{(i)}(\seq{v}_i))_{i \in \mathcal{I}_K}, Y^n) \not \in T_{n,\gamma}(\Phi_n, \seq{v}_{\mathcal{I}_K})\} \IEEEnonumber \\
&+\: \sum_{\scriptstyle A \subseteq \mathcal{I}_K, \atop \scriptstyle A \ne \emptyset} \sum_{\scriptstyle \hat{\seq{v}}_{\mathcal{I}_K}: \hat{\seq{v}}_{A^c} = \seq{v}_{A^c}, \atop \scriptstyle (\hat{\seq{v}}_i \ne \seq{v}_i)_{i \in A}} \Pr\{((\Phi_n^{(i)}(\hat{\seq{v}}_i))_{i \in \mathcal{I}_K}, Y^n) \IEEEnonumber \\
&\in T_{n,\gamma}^A(\Phi_n, \hat{\seq{v}}_{\mathcal{I}_K})\} \label{eq:UBofEpsilonV}
\end{IEEEeqnarray}
where $Y^n$ denotes the channel output corresponding to the input $(\Phi_n^{(i)}(\seq{v}_i))_{i \in \mathcal{I}_K}$. Since the first term on the right-hand side of \eqref{eq:UBofEpsilonV} can be written as
\begin{IEEEeqnarray*}{Cl}
&B_n(\seq{v}_{\mathcal{I}_K}) \\
\eqdef &\Pr\{((\Phi_n^{(i)}(\seq{v}_i))_{i \in \mathcal{I}_K}, Y^n) \not \in T_{n,\gamma}(\Phi_n, \seq{v}_{\mathcal{I}_K})\} \\
= &\sum_{(\seq{x}_{\mathcal{I}_K}, \seq{y}) \not \in T_{n,\gamma}(\Phi_n, \seq{v}_{\mathcal{I}_K})} W^n(\seq{y}|\seq{x}_{\mathcal{I}_K}) \\
&\Pr\{(\Phi_n^{(i)}(\seq{v}_i))_{i \in \mathcal{I}_K} = \seq{x}_{\mathcal{I}_K}\} \\
\eqvar{(a)} &\sum_{(\seq{x}_{\mathcal{I}_K}, \seq{y}) \not \in T_{n,\gamma}(\Phi_n, \seq{v}_{\mathcal{I}_K})} W^n(\seq{y}|\seq{x}_{\mathcal{I}_K}) \prod_{i \in \mathcal{I}_K} P_{X_i^n|V_i^n}(\seq{x}_i|\seq{v}_i) \\
= &\sum_{(\seq{x}_{\mathcal{I}_K}, \seq{y}) \not \in T_{n,\gamma}(\Phi_n, \seq{v}_{\mathcal{I}_K})} P_{X_{\mathcal{I}_K}^nY^n|V_{\mathcal{I}_K}^n}(\seq{x}_{\mathcal{I}_K},\seq{y}|\seq{v}_{\mathcal{I}_K})
\end{IEEEeqnarray*}
where (a) follows from \eqref{eq:Condition1OfEncoders} and \eqref{eq:ConditionalDistributionX|V}, it follows that
\begin{IEEEeqnarray}{Cl}
&\sum_{\seq{v}_{\mathcal{I}_K} \in \prod_{i \in \mathcal{I}_K} \mathcal{V}_i^n} P_{V_{\mathcal{I}_K}^n}(\seq{v}_{\mathcal{I}_K}) B_n(\seq{v}_{\mathcal{I}_K}) \IEEEnonumber \\
= &\sum_{\seq{v}_{\mathcal{I}_K} \in \prod_{i \in \mathcal{I}_K} \mathcal{V}_i^n} P_{V_{\mathcal{I}_K}^n}(\seq{v}_{\mathcal{I}_K}) \sum_{(\seq{x}_{\mathcal{I}_K}, \seq{y}) \not \in T_{n,\gamma}(\Phi_n, \seq{v}_{\mathcal{I}_K})} \IEEEnonumber \\
&P_{X_{\mathcal{I}_K}^nY^n|V_{\mathcal{I}_K}^n}(\seq{x}_{\mathcal{I}_K},\seq{y}|\seq{v}_{\mathcal{I}_K}) \IEEEnonumber \\
= &\sum_{(\seq{v}_{\mathcal{I}_K},\seq{x}_{\mathcal{I}_K},\seq{y}) \not \in T_{n,\gamma}(\Phi_n)} P_{V_{\mathcal{I}_K}^nX_{\mathcal{I}_K}^nY^n}(\seq{v}_{\mathcal{I}_K}, \seq{x}_{\mathcal{I}_K}, \seq{y}) \IEEEnonumber \\
= &\Pr\{(V_{\mathcal{I}_K}^n, X_{\mathcal{I}_K}^n, Y^n) \not \in T_{n,\gamma}(\Phi_n)\}. \label{eq:UBofAverageAn}
\end{IEEEeqnarray}
On the other hand, the second term on the right-hand side of \eqref{eq:UBofEpsilonV} can be written as
$$
C_n(\seq{v}_{\mathcal{I}_K}) = \sum_{\scriptstyle A \subseteq \mathcal{I}_K, \atop \scriptstyle A \ne \emptyset} C_n(A, \seq{v}_{\mathcal{I}_K})
$$
and
\begin{IEEEeqnarray*}{Cl}
&C_n(A, \seq{v}_{\mathcal{I}_K}) \\
\eqdef &\sum_{\scriptstyle \hat{\seq{v}}_{\mathcal{I}_K}: \hat{\seq{v}}_{A^c} = \seq{v}_{A^c}, \atop \scriptstyle (\hat{\seq{v}}_i \ne \seq{v}_i)_{i \in A}} \compact\compact\compact \Pr\{((\Phi_n^{(i)}(\hat{\seq{v}}_i))_{i \in \mathcal{I}_K}, Y^n) \in T_{n,\gamma}^A(\Phi_n, \hat{\seq{v}}_{\mathcal{I}_K})\} \\
= &\sum_{\scriptstyle \hat{\seq{v}}_{\mathcal{I}_K}: \hat{\seq{v}}_{A^c} = \seq{v}_{A^c}, \atop \scriptstyle (\hat{\seq{v}}_i \ne \seq{v}_i)_{i \in A}} \sum_{(\seq{x}_{\mathcal{I}_K}, \seq{y}) \in \prod_{i \in \mathcal{I}_K} \mathcal{X}_i^n \times \mathcal{Y}^n} W^n(\seq{y}|\seq{x}_{\mathcal{I}_K}) \\
&\Pr\{(\Phi_n^{(i)}(\seq{v}_i))_{i \in \mathcal{I}_K} = \seq{x}_{\mathcal{I}_K}\} \sum_{(\hat{\seq{x}}_{\mathcal{I}_K}, \seq{y}) \in T_{n,\gamma}^A(\Phi_n, \hat{\seq{v}}_{\mathcal{I}_K})} \\
&\Pr\{(\Phi_n^{(i)}(\hat{\seq{v}}_i))_{i \in \mathcal{I}_K} = \hat{\seq{x}}_{\mathcal{I}_K} | (\Phi_n^{(i)}(\seq{v}_i))_{i \in \mathcal{I}_K} = \seq{x}_{\mathcal{I}_K}\} \\
\eqvar{(a)} &\sum_{\scriptstyle \hat{\seq{v}}_{\mathcal{I}_K}: \hat{\seq{v}}_{A^c} = \seq{v}_{A^c}, \atop \scriptstyle (\hat{\seq{v}}_i \ne \seq{v}_i)_{i \in A}} \sum_{(\seq{x}_{\mathcal{I}_K}, \seq{y}) \in \prod_{i \in \mathcal{I}_K} \mathcal{X}_i^n \times \mathcal{Y}^n} W^n(\seq{y}|\seq{x}_{\mathcal{I}_K}) \\
&\prod_{i \in \mathcal{I}_K} P_{X_i^n|V_i^n}(\seq{x}_i|\seq{v}_i) \sum_{\scriptstyle (\hat{\seq{x}}_{\mathcal{I}_K}, \seq{y}) \in T_{n,\gamma}^A(\Phi_n, \hat{\seq{v}}_{\mathcal{I}_K}), \atop \scriptstyle \hat{\seq{x}}_{A^c} = \seq{x}_{A^c}} \\
&\Pr\{(\Phi_n^{(i)}(\hat{\seq{v}}_i))_{i \in A} = \hat{\seq{x}}_A | (\Phi_n^{(i)}(\seq{v}_i))_{i \in \mathcal{I}_K} = \seq{x}_{\mathcal{I}_K}\} \\
\levar{(b)} &\sum_{\scriptstyle \hat{\seq{v}}_{\mathcal{I}_K}: \hat{\seq{v}}_{A^c} = \seq{v}_{A^c}, \atop \scriptstyle (\hat{\seq{v}}_i \ne \seq{v}_i)_{i \in A}} \sum_{(\seq{x}_{\mathcal{I}_K}, \seq{y}) \in \prod_{i \in \mathcal{I}_K} \mathcal{X}_i^n \times \mathcal{Y}^n} \\
&P_{X_{\mathcal{I}_K}^nY^n|V_{\mathcal{I}_K}^n}(\seq{x}_{\mathcal{I}_K},\seq{y}|\seq{v}_{\mathcal{I}_K}) \sum_{\scriptstyle (\hat{\seq{x}}_{\mathcal{I}_K}, \seq{y}) \in T_{n,\gamma}^A(\Phi_n, \hat{\seq{v}}_{\mathcal{I}_K}), \atop \scriptstyle \hat{\seq{x}}_{A^c} = \seq{x}_{A^c}} \\
&e^{n\rho(\Phi_n^{(A)})(\hat{\seq{v}}_A, \hat{\seq{x}}_A)} \prod_{i \in A} P_{X_i^n|V_i^n}(\hat{\seq{x}}_i | \hat{\seq{v}}_i) \\
\le &\sum_{\hat{\seq{v}}_{\mathcal{I}_K}: \hat{\seq{v}}_{A^c} = \seq{v}_{A^c}} \sum_{(\seq{x}_{A^c}, \seq{y}) \in \prod_{i \in A^c} \mathcal{X}_i^n \times \mathcal{Y}^n} \\
&\sum_{\scriptstyle (\hat{\seq{x}}_{\mathcal{I}_K}, \seq{y}) \in T_{n,\gamma}^A(\Phi_n, \hat{\seq{v}}_{\mathcal{I}_K}), \atop \scriptstyle \hat{\seq{x}}_{A^c} = \seq{x}_{A^c}} e^{n\rho(\Phi_n^{(A)})(\hat{\seq{v}}_A, \hat{\seq{x}}_A)} \\
&P_{X_{A^c}^nY^n|V_{\mathcal{I}_K}^n}(\seq{x}_{A^c}, \seq{y}|\seq{v}_{\mathcal{I}_K}) \prod_{i \in A} P_{X_i^n|V_i^n}(\hat{\seq{x}}_i | \hat{\seq{v}}_i)
\end{IEEEeqnarray*}
where (a) follows from \eqref{eq:Condition1OfEncoders}, \eqref{eq:ConditionalDistributionX|V}, and the fact that $\Phi_n^{(i)}(\hat{\seq{v}}_i) = \Phi_n^{(i)}(\seq{v}_i)$ for all $i \in A^c$ since $\hat{\seq{v}}_{A^c} = \seq{v}_{A^c}$, and (b) follows from \eqref{eq:Condition2OfEncoders} and \eqref{eq:DistributionOfVXY}. Therefore, it follows that
\begin{IEEEeqnarray}{Cl}
&\sum_{\seq{v}_{\mathcal{I}_K} \in \prod_{i \in \mathcal{I}_K} \mathcal{V}_i^n} P_{V_{\mathcal{I}_K}^n}(\seq{v}_{\mathcal{I}_K}) C_n(\seq{v}_{\mathcal{I}_K}) \IEEEnonumber \\
\le &\sum_{\scriptstyle A \subseteq \mathcal{I}_K, \atop \scriptstyle A \ne \emptyset} \sum_{\seq{v}_{A^c} \in \prod_{i \in A^c} \mathcal{V}_i^n} \sum_{\hat{\seq{v}}_{\mathcal{I}_K}: \hat{\seq{v}}_{A^c} = \seq{v}_{A^c}} \sum_{(\seq{x}_{A^c}, \seq{y}) \in \prod_{i \in A^c} \mathcal{X}_i^n \times \mathcal{Y}^n} \IEEEnonumber \\
&\sum_{\scriptstyle (\hat{\seq{x}}_{\mathcal{I}_K}, \seq{y}) \in T_{n,\gamma}^A(\Phi_n, \hat{\seq{v}}_{\mathcal{I}_K}), \atop \scriptstyle \hat{\seq{x}}_{A^c} = \seq{x}_{A^c}} \compact\compact\compact e^{n\rho(\Phi_n^{(A)})(\hat{\seq{v}}_A, \hat{\seq{x}}_A)} \prod_{i \in A} P_{X_i^n|V_i^n}(\hat{\seq{x}}_i | \hat{\seq{v}}_i) \IEEEnonumber \\
&\sum_{\seq{v}_A \in \prod_{i \in A} \mathcal{V}_i^n} P_{V_{\mathcal{I}_K}^n}(\seq{v}_{\mathcal{I}_K}) P_{X_{A^c}^nY^n|V_{\mathcal{I}_K}^n}(\seq{x}_{A^c}, \seq{y}|\seq{v}_{\mathcal{I}_K}) \IEEEnonumber \\
= &\sum_{\scriptstyle A \subseteq \mathcal{I}_K, \atop \scriptstyle A \ne \emptyset} \sum_{\hat{\seq{v}}_{\mathcal{I}_K} \in \prod_{i \in \mathcal{I}_K} \mathcal{V}_i^n} \sum_{(\hat{\seq{x}}_{\mathcal{I}_K}, \seq{y}) \in T_{n,\gamma}^A(\Phi_n, \hat{\seq{v}}_{\mathcal{I}_K})} \IEEEnonumber \\
&e^{n\rho(\Phi_n^{(A)})(\hat{\seq{v}}_A, \hat{\seq{x}}_A)} \prod_{i \in A} P_{X_i^n|V_i^n}(\hat{\seq{x}}_i | \hat{\seq{v}}_i) \IEEEnonumber \\
&P_{V_{A^c}^nX_{A^c}^nY^n}(\hat{\seq{v}}_{A^c}, \hat{\seq{x}}_{A^c}, \seq{y}) \IEEEnonumber \\
\levar{(a)} &e^{-n\gamma} \sum_{\scriptstyle A \subseteq \mathcal{I}_K, \atop \scriptstyle A \ne \emptyset} \sum_{\hat{\seq{v}}_{\mathcal{I}_K} \in \prod_{i \in \mathcal{I}_K} \mathcal{V}_i^n} \sum_{(\hat{\seq{x}}_{\mathcal{I}_K}, \seq{y}) \in T_{n,\gamma}^A(\Phi_n, \hat{\seq{v}}_{\mathcal{I}_K})} \IEEEnonumber \\
&W^n(\seq{y}|\hat{\seq{x}}_{\mathcal{I}_K}) P_{V_A^n|V_{A^c}^n}(\hat{\seq{v}}_A | \hat{\seq{v}}_{A^c}) \prod_{i \in A} P_{X_i^n|V_i^n}(\hat{\seq{x}}_i | \hat{\seq{v}}_i) \IEEEnonumber \\
&P_{V_{A^c}^nX_{A^c}^n}(\hat{\seq{v}}_{A^c}, \hat{\seq{x}}_{A^c}) \IEEEnonumber \\
\le &e^{-n\gamma} \sum_{\scriptstyle A \subseteq \mathcal{I}_K, \atop \scriptstyle A \ne \emptyset} \sum_{\hat{\seq{v}}_{\mathcal{I}_K} \in \prod_{i \in \mathcal{I}_K} \mathcal{V}_i^n} \sum_{\hat{\seq{x}}_{\mathcal{I}_K} \in \prod_{i \in \mathcal{I}_K} \mathcal{X}_i^n} \sum_{\seq{y} \in \mathcal{Y}^n} \IEEEnonumber \\
&P_{V_{\mathcal{I}_K}^nX_{\mathcal{I}_K}^nY^n}(\hat{\seq{v}}_{\mathcal{I}_K}, \hat{\seq{x}}_{\mathcal{I}_K}, \seq{y}) \IEEEnonumber \\
= &(2^K - 1)e^{-n\gamma},\label{eq:UBofAverageBn}
\end{IEEEeqnarray}
where (a) follows from the inequality
\begin{IEEEeqnarray*}{Cl}
&e^{n\rho(\Phi_n^{(A)})(\hat{\seq{v}}_A, \hat{\seq{x}}_A)} P_{Y^n|X_{A^c}^nV_{A^c}^n}(\seq{y} | \hat{\seq{x}}_{A^c}, \hat{\seq{v}}_{A^c}) \\
\le &e^{-n\gamma} P_{V_A^n|V_{A^c}^n}(\hat{\seq{v}}_A | \hat{\seq{v}}_{A^c}) W^n(\seq{y}|\hat{\seq{x}}_{\mathcal{I}_K})
\end{IEEEeqnarray*}
implied by \eqref{eq:GoodSet2ForDecoding}. Hence from \eqref{eq:AverageErrorFormula}, \eqref{eq:UBofEpsilonV}, \eqref{eq:UBofAverageAn}, and \eqref{eq:UBofAverageBn}, it follows that
\begin{IEEEeqnarray*}{rCl}
\epsilon_n &= &\sum_{\seq{v}_{\mathcal{I}_K} \in \prod_{i \in \mathcal{I}_K} \mathcal{V}_i^n} P_{V_{\mathcal{I}_K}^n}(\seq{v}_{\mathcal{I}_K}) \epsilon_n(\seq{v}_{\mathcal{I}_K}) \\
&\le &\sum_{\seq{v}_{\mathcal{I}_K} \in \prod_{i \in \mathcal{I}_K} \mathcal{V}_i^n} P_{V_{\mathcal{I}_K}^n}(\seq{v}_{\mathcal{I}_K}) B_n(\seq{v}_{\mathcal{I}_K}) \\
& &+\: \sum_{\seq{v}_{\mathcal{I}_K} \in \prod_{i \in \mathcal{I}_K} \mathcal{V}_i^n} P_{V_{\mathcal{I}_K}^n}(\seq{v}_{\mathcal{I}_K}) C_n(\seq{v}_{\mathcal{I}_K}) \\
&\le &\Pr\{(V_{\mathcal{I}_K}^n, X_{\mathcal{I}_K}^n, Y^n) \not \in T_{n,\gamma}(\Phi_n)\} + (2^K - 1)e^{-n\gamma}.
\end{IEEEeqnarray*}
This completes the proof.
\end{proofof}

\begin{proofof}{Corollary \ref{co:GeneralizedDirectPartForMAC}}
For all nonempty set $A \subseteq \mathcal{I}_K$, we have
\begin{IEEEeqnarray*}{Cl}
&\plimsup_{n \to \infty} (h(V_A^n|V_{A^c}^n) - i(X_A^n; Y^n|X_{A^c}^n, V_{A^c}^n) \\
&+\: \rho(\Phi_n^{(A)})(V_A^n, X_A^n)) \\
\levar{(a)} &\plimsup_{n \to \infty} h(V_A^n|V_{A^c}^n) - \pliminf_{n \to \infty} i(X_A^n; Y^n|X_{A^c}^n, V_{A^c}^n) \\
&+\: \plimsup_{n \to \infty} \rho(\Phi_n^{(A)})(V_A^n, X_A^n) \\
= &\overline{H}(\bm{V}_A | \bm{V}_{A^c}) - \underline{I}(\bm{X}_A; \bm{Y}|\bm{X}_{A^c}, \bm{V}_{A^c}) \\
&+\: \plimsup_{n \to \infty} \rho(\Phi_n^{(A)})(V_A^n, X_A^n) \\
\ltvar{(b)} &0
\end{IEEEeqnarray*}
where (a) follows from the properties that
$$
\plimsup_{n \to \infty} (X_n + Y_n) \le \plimsup_{n \to \infty} X_n + \plimsup_{n \to \infty} Y_n
$$
$$
\pliminf_{n \to \infty} X_n = -\plimsup_{n \to \infty} (-X_n),
$$
and (b) follows from the condition \eqref{eq:AchievableCondition}. Then, by the definition of limit superior in probability and \eqref{eq:GoodSet2ForDecoding}, we obtain that
$$
\lim_{n \to \infty} \Pr\{(V_{\mathcal{I}_K}^n, X_{\mathcal{I}_K}^n, Y^n) \not \in T_{n, \gamma_A}^A(\Phi_n)\} = 0
$$
for some $\gamma_A > 0$. Furthermore, by the definition \eqref{eq:GoodSetForDecoding}, we have
\begin{IEEEeqnarray*}{Cl}
&\lim_{n \to \infty} \Pr\{(V_{\mathcal{I}_K}^n, X_{\mathcal{I}_K}^n, Y^n) \not \in T_{n, \gamma}(\Phi_n)\} \\
\le &\lim_{n \to \infty} \sum_{A \subseteq \mathcal{I}_K, A \ne \emptyset} \Pr\{(V_{\mathcal{I}_K}^n, X_{\mathcal{I}_K}^n, Y^n) \not \in T_{n, \gamma}^A(\Phi_n)\} \\
\eqvar{(a)} &0
\end{IEEEeqnarray*}
where (a) follows from the fact that $\mathcal{I}_K$ is a finite set, and
$$
\gamma = \min_{A \subseteq \mathcal{I}_K, A \ne \emptyset} \gamma_A > 0.
$$
Then, from Theorem \ref{th:GeneralizedDirectPartForMAC}, it follows that there exits a code such that $\bm{V}_{\mathcal{I}_K}$ is transmissible over $\bm{W}$.
\end{proofof}

\begin{proofof}{Corollary \ref{co:DirectPartForMemorylessMAC}}
By regarding every $N$ consecutive symbols in a sequence as a single symbol, we may define a new source
$$
\dot{\bm{V}}_{\mathcal{I}_K} = \{\dot{V}_{\mathcal{I}_K}^n\}_{n=1}^\infty = \{V_{\mathcal{I}_K}^{nN}\}_{n=1}^\infty
$$
and a new channel
$$
\dot{\bm{W}} = \{\dot{W}^n\}_{n=1}^\infty = \{W^{nN}\}_{n=1}^\infty.
$$
Furthermore, let
$$
\dot{\bm{X}}_{\mathcal{I}_K} = \{\dot{X}_{\mathcal{I}_K}^n\}_{n=1}^\infty = \{X_{\mathcal{I}_K}^{nN}\}_{n=1}^\infty
$$
and
$$
\dot{\bm{Y}} = \{\dot{Y}^n\}_{n=1}^\infty = \{Y^{nN}\}_{n=1}^\infty,
$$
then it follows from \eqref{eq:ConditionalDistributionX|V2} that the triple $(\dot{\bm{V}}_{\mathcal{I}_K}, \dot{\bm{X}}_{\mathcal{I}_K}, \dot{\bm{Y}})$ are jointly stationary and memoryless. Hence, we have
$$
\overline{H}(\dot{\bm{V}}_A|\dot{\bm{V}}_{A^c}) = H(V_A^N|V_{A^c}^N) = N H(V_A|V_{A^c})
$$
and
$$
\underline{I}(\dot{\bm{X}}_A; \dot{\bm{Y}}|\dot{\bm{X}}_{A^c}, \dot{\bm{V}}_{A^c}) = I(X_A^N; Y^N|X_{A^c}^N, V_{A^c}^N)
$$
for all nonempty set $A \subseteq \mathcal{I}_K$. The corollary then follows from \eqref{eq:AchievableCondition2} and Corollary \ref{co:GeneralizedDirectPartForMAC}.
\end{proofof}

\begin{proofof}{Lemma \ref{le:IndependenceLC2}}
From the definition \eqref{eq:DefinitionOfGoodEncoderBasedOnLC}, it follows that
\begin{IEEEeqnarray*}{rCl}
\Pr\{\Phi_{F_n, q_n}(\seq{v}) = \seq{x}\} &= &\sum_{\seq{u} \in q_n^{-1}(\seq{v}, \seq{x})} \Pr\{\hat{F}_n(\seq{v}) = \seq{u}\} \\
&\eqvar{(a)} &\sum_{\seq{u} \in q_n^{-1}(\seq{v}, \seq{x})} |\mathcal{U}|^{-l} \\
&\eqvar{(b)} &P_{X^m|V^n}(\seq{x}|\seq{v})
\end{IEEEeqnarray*}
where (a) follows from Proposition \ref{pr:IndependenceLC1} and (b) follows from the definition \eqref{eq:SimulationOfConditionalProbability}. Again from \eqref{eq:DefinitionOfGoodEncoderBasedOnLC}, it follows that
\begin{IEEEeqnarray*}{Cl}
&\Pr\{\Phi_{F_n, q_n}(\seq{v}) = \seq{x}, \Phi_{F_n, q_n}(\hat{\seq{v}}) = \hat{\seq{x}}\} \\
= &\sum_{\scriptstyle \seq{u} \in q_n^{-1}(\seq{v}, \seq{x}), \atop \scriptstyle \hat{\seq{u}} \in q_n^{-1}(\hat{\seq{v}}, \hat{\seq{x}})} \Pr\{\hat{F}(\seq{v}) = \seq{u}, \hat{F}(\hat{\seq{v}}) = \hat{\seq{u}} \} \\
\eqvar{(a)} &\sum_{\scriptstyle \seq{u} \in q_n^{-1}(\seq{v}, \seq{x}), \atop \scriptstyle \hat{\seq{u}} \in q_n^{-1}(\hat{\seq{v}}, \hat{\seq{x}})} |\mathcal{U}|^{-2l} \alpha(F_n)(P_{\hat{\seq{v}}-\seq{v}}, P_{\hat{\seq{u}}-\seq{u}}) \\
\eqvar{(b)} &\frac{P_{X^m|V^n}(\seq{x}|\seq{v}) P_{X^m|V^n}(\hat{\seq{x}}|\hat{\seq{v}})}{|q_n^{-1}(\seq{v}, \seq{x})| |q_n^{-1}(\hat{\seq{v}}, \hat{\seq{x}})|} \\
&\sum_{\scriptstyle \seq{u} \in q_n^{-1}(\seq{v}, \seq{x}), \atop \scriptstyle \hat{\seq{u}} \in q_n^{-1}(\hat{\seq{v}}, \hat{\seq{x}})} \alpha(F_n)(P_{\hat{\seq{v}}-\seq{v}}, P_{\hat{\seq{u}}-\seq{u}}) \\
\eqvar{(c)} &\beta(F_n, q_n)(\seq{v}, \hat{\seq{v}}, \seq{x}, \hat{\seq{x}}) P_{X^m|V^n}(\seq{x}|\seq{v}) P_{X^m|V^n}(\hat{\seq{x}}|\hat{\seq{v}})
\end{IEEEeqnarray*}
where (a) follows from Proposition \ref{pr:IndependenceLC1}, (b) follows from \eqref{eq:SimulationOfConditionalProbability}, and (c) follows from \eqref{eq:DefinitionOfBeta}. This concludes the equality \eqref{eq:Independence1ofLC2}, and \eqref{eq:Independence2ofLC2} is an easy consequence of \eqref{eq:Independence1ofLC2}.
\end{proofof}

\subsection{Proofs of Results in Section \ref{sec:Discussion}}\label{subsec:ProofOfSectionIV}

\begin{proofof}{Proposition \ref{pr:RelationBetweenRLC&SLC}}
If $F_n$ is good, we have
$$
E \biggl[ \frac{S_{\mathcal{X}\mathcal{Y}}(F)(P, Q)}{S_{\mathcal{X}\mathcal{Y}}(\mathcal{X}^n \times \mathcal{Y}^{m_n})(P, Q)} \biggr] \le 1
$$
for any $P \in \mathcal{P}_n(\mathcal{X}) \backslash \{P_{0^n}\}$ and $Q \in \mathcal{P}_{m_n}(\mathcal{Y})$. Then by Markov's inequality, we have
\begin{IEEEeqnarray*}{Cl}
&\Pr \biggl\{ \frac{S_{\mathcal{X}\mathcal{Y}}(F)(P, Q)}{S_{\mathcal{X}\mathcal{Y}}(\mathcal{X}^n \times \mathcal{Y}^{m_n})(P, Q)} \ge (n+1)^{c_1} (m_n+1)^{c_2} \biggr\} \\
\le &\frac{1}{(n+1)^{c_1} (m_n+1)^{c_2}} E\biggl[ \frac{S_{\mathcal{X}\mathcal{Y}}(F)(P, Q)}{S_{\mathcal{X}\mathcal{Y}}(\mathcal{X}^n \times \mathcal{Y}^{m_n})(P, Q)} \biggr] \\
\le &\frac{1}{(n+1)^{c_1} (m_n+1)^{c_2}}
\end{IEEEeqnarray*}
for all $P \in \mathcal{P}_n(\mathcal{X}) \backslash P_{0^n}$ and $Q \in \mathcal{P}_{m_n}(\mathcal{Y})$. And furthermore, we have
\begin{IEEEeqnarray*}{Cl}
&\Pr \biggl\{ \max_{\scriptstyle P \in \mathcal{P}_{n}(\mathcal{X}) \backslash \{P_{0^n}\}, \atop \scriptstyle Q \in \mathcal{P}_{m_n}(\mathcal{Y})} \frac{S_{\mathcal{X}\mathcal{Y}}(F)(P, Q)}{S_{\mathcal{X}\mathcal{Y}}(\mathcal{X}^n \times \mathcal{Y}^{m_n})(P, Q)} \\
&\ge (n+1)^{c_1} (m_n+1)^{c_2} \biggr\} \\
\le &\sum_{\scriptstyle P \in \mathcal{P}_{n}(\mathcal{X}) \backslash \{P_{0^n}\}, \atop \scriptstyle Q \in \mathcal{P}_{m_n}(\mathcal{Y})} \Pr \biggl\{ \frac{S_{\mathcal{X}\mathcal{Y}}(F)(P, Q)}{S_{\mathcal{X}\mathcal{Y}}(\mathcal{X}^n \times \mathcal{Y}^{m_n})(P, Q)} \\
&\ge (n+1)^{c_1} (m_n+1)^{c_2} \biggr\} \\
\le &\frac{|\mathcal{P}_{n}(\mathcal{X})| |\mathcal{P}_{m_n}(\mathcal{Y})|}{(n+1)^{c_1} (m_n+1)^{c_2}} \\
\levar{(a)} &\frac{1}{(n+1)^{c_1 - |\mathcal{X}|} (m_n+1)^{c_2 - |\mathcal{Y}|}}
\end{IEEEeqnarray*}
where (a) follows the basic fact $|\mathcal{P}_n(\mathcal{X})| \le (n+1)^{|\mathcal{X}|}$ in the method of types \cite{JSCC:Csiszar198100}. This together with the conditions $c_1 > |\mathcal{X}|$ and $c_2 > |\mathcal{Y}|$ then concludes \eqref{eq:RelationBetwwenRLC&SLC}.

By analogous arguments, we can prove that for any $\delta$-asymptotically good random linear code, there exists at least one sample code which is $\delta$-asymptotically good.
\end{proofof}

\begin{proofof}{Theorem \ref{th:DistanceProperty}}
According to the definition \eqref{eq:DefinitionOfB}, we have
\begin{IEEEeqnarray*}{Cl}
&E[|B(F_n, h_{\mathcal{X}}, h_{\mathcal{Y}})|] \\
= &E\biggl[\sum_{\scriptstyle \seq{x}: \seq{x} \ne 0^n, H(P_{\seq{x}}) \le h_{\mathcal{X}}, \atop \scriptstyle \seq{y}: H(P_{\seq{y}}) \le h_{\mathcal{Y}}} 1\{F_n(\seq{x}) = \seq{y}\}\biggr] \\
= &\sum_{\scriptstyle \seq{x}: \seq{x} \ne 0^n, H(P_{\seq{x}}) \le h_{\mathcal{X}}, \atop \scriptstyle \seq{y}: H(P_{\seq{y}}) \le h_{\mathcal{Y}}} \Pr\{F_n(\seq{x}) = \seq{y}\}.
\end{IEEEeqnarray*}
And for any $\epsilon > 0$ , we further have
\begin{IEEEeqnarray}{Cl}
&E[|B(F_n, h_{\mathcal{X}}, h_{\mathcal{Y}})|] \IEEEnonumber \\
\eqvar{(a)} &\sum_{\scriptstyle \seq{x}: \seq{x} \ne 0^n, H(P_{\seq{x}}) \le h_{\mathcal{X}}, \atop \scriptstyle \seq{y}: H(P_{\seq{y}}) \le h_{\mathcal{Y}}} |\mathcal{Y}|^{-m_n} \alpha(F_n)(P_{\seq{x}}, P_{\seq{y}}) \IEEEnonumber \\
\levar{(b)} &\sum_{\scriptstyle \seq{x}: \seq{x} \ne 0^n, H(P_{\seq{x}}) \le h_{\mathcal{X}}, \atop \scriptstyle \seq{y}: H(P_{\seq{y}}) \le h_{\mathcal{Y}}} |\mathcal{Y}|^{-m_n} e^{n(\delta + \epsilon_1)} \IEEEnonumber \\
\levar{(c)} &|\mathcal{Y}|^{-m_n} e^{n(\delta + \epsilon_1)} e^{n(h_{\mathcal{X}} + \epsilon_2)} e^{m_n(h_{\mathcal{Y}} + \epsilon_3)} \IEEEnonumber \\
\levar{(d)} &\exp\{-m_n(\ln|\mathcal{Y}| - (h_{\mathcal{X}} + \delta) \overline{R}(\bm{F}) - h_{\mathcal{Y}} - \epsilon)\} \label{eq:UpperBoundOfCountsOfB}
\end{IEEEeqnarray}
for sufficiently large $n$, where (a) follows from Proposition \ref{pr:SpectrumPropertyOfFunctions}, (b) from the definition \eqref{eq:ConditionOfAsymptoticGoodLinearCodes} of a $\delta$-asymptotically good linear code, (c) from the inequality
$$
\biggl| \bigcup_{H(P) \le h} \mathcal{T}_{P}^n(\mathcal{X}) \biggr| \le e^{n(h + \epsilon)}
$$
in the method of types \cite{JSCC:Csiszar198100}, and (d) follows from the condition \eqref{eq:RateOfLinearCode}. Then the upper bound \eqref{eq:UpperBoundOfCountsOfB} together with the condition \eqref{eq:DistanceInequality} concludes the limit \eqref{eq:DistanceProperty}, because the number $\epsilon$ in \eqref{eq:UpperBoundOfCountsOfB} can be arbitrarily small.

Next, for any $\gamma > 0$, we have
\begin{IEEEeqnarray*}{Cl}
&\Pr\bigl\{\min_{\seq{x} \ne 0^n} [(H(P_{\seq{x}}) + \delta) \overline{R}(\bm{F}) + H(P_{F_n(\seq{x})})] < \ln|\mathcal{Y}| - \gamma \bigr\} \\
= &E\bigl[ 1\{\min_{\seq{x} \ne 0^n} [(H(P_{\seq{x}}) + \delta) \overline{R}(\bm{F}) + H(P_{F_n(\seq{x})})] \\
&< \ln|\mathcal{Y}| - \gamma\} \bigr] \\
\le &E\biggl[ \sum_{\scriptstyle P \in \mathcal{P}_n(\mathcal{X}) \backslash \{P_{0^n}\}, \atop \scriptstyle Q \in \mathcal{P}_{m_n}(\mathcal{Y})} \sum_{\scriptstyle \seq{x} \in \mathcal{T}_P^n(\mathcal{X}), \atop \scriptstyle F_n(\seq{x}) \in \mathcal{T}_Q^{m_n}(\mathcal{Y})} \\
&1\{(H(P) + \delta) \overline{R}(\bm{F}) + H(Q) < \ln|\mathcal{Y}| - \gamma\} \biggr] \\
= &\sum_{\scriptstyle P \in \mathcal{P}_n(\mathcal{X}) \backslash \{P_{0^n}\}, \atop \scriptstyle Q \in \mathcal{P}_{m_n}(\mathcal{Y})} E \biggl[ \sum_{\scriptstyle \seq{x} \in \mathcal{T}_P^n(\mathcal{X}), \atop \scriptstyle F_n(\seq{x}) \in \mathcal{T}_Q^{m_n}(\mathcal{Y})} \\
&1\{ H(Q) < \ln|\mathcal{Y}| - (H(P) + \delta) \overline{R}(\bm{F}) - \gamma\} \biggr] \\
\levar{(a)} &\sum_{\scriptstyle P \in \mathcal{P}_n(\mathcal{X}) \backslash \{P_{0^n}\}, \atop \scriptstyle Q \in \mathcal{P}_{m_n}(\mathcal{Y})} E \biggl[ \bigl| B(F_n, H(P), \\
&\ln|\mathcal{Y}| - (H(P) + \delta) \overline{R}(\bm{F}) - \gamma) \bigr| \biggr] \\
\levar{(b)} &(n + 1)^{|\mathcal{X}|} (m_n + 1)^{|\mathcal{Y}|} \exp\{-m_n(\gamma - \epsilon)\} \\
\to &0
\end{IEEEeqnarray*}
as $n \to \infty$, where (a) follows from the definition \eqref{eq:DefinitionOfB}, and (b) from \eqref{eq:UpperBoundOfCountsOfB} and the inequality $|\mathcal{P}_n(\mathcal{X})| \le (n+1)^{|\mathcal{X}|}$ concluded by the method of types \cite{JSCC:Csiszar198100}. This together with the definition \eqref{eq:DefinitionOfPLiminf} then concludes the second inequality of \eqref{eq:DistanceProperty2}, and the first inequality follows easily from the simple inequality $H(P_{\seq{x}}) \le \ln |\mathcal{X}|$ for all $\seq{x} \in \mathcal{X}^n$.
\end{proofof}

\begin{proofof}{Corollary \ref{co:SystematicCode}}
Let $a$ be a nonzero element in $\mathcal{X}$. Because $\{F_n\}_{n=1}^\infty$ is asymptotically good, we then have
\begin{IEEEeqnarray}{Cl}
&\pliminf_{n \to \infty} H(P_{F_n(a^n)}) \IEEEnonumber \\
\ge &\pliminf_{n \to \infty} \min_{\seq{x} \ne 0^n} [H(P_{\seq{x}}) \overline{R}(\bm{F}) + H(P_{F_n(\seq{x})})] \IEEEnonumber \\
\ge &\ln|\mathcal{X}| \label{eq:SystemticCode}
\end{IEEEeqnarray}
by Theorem \ref{th:DistanceProperty}. If we assume that $\overline{R}(\bm{F}) > 1/|\mathcal{X}|$, then there exists at least a subsequence $\{F_{n_l}\}_{l=1}^\infty$ of codes in $\{F_n\}_{n=1}^\infty$ such that
$$
R(F_{n_l}) > \frac{1}{|\mathcal{X}|} + \epsilon_1
$$
for some positive real number $\epsilon_1$ and sufficiently large $l$. Since the codes are systematic, we have
\begin{equation}\label{eq:ProbabilityOfNonzeroSymbol}
P_{F_{n_l}(a^{n_l})}(a) > \frac{1}{|\mathcal{X}|} + \epsilon_1
\end{equation}
for sufficiently large $l$. The entropy $H(P_{F_{n_l}(a^{n_l})})$ can be bounded above by
$$
h(P_{F_{n_l}(a^{n_l})}(a)) + (1 - P_{F_{n_l}(a^{n_l})}(a)) \ln (|\mathcal{X}| - 1)
$$
where $h(x)= - x \ln x - (1 - x) \ln (1 - x)$. Define
$$
g(x) = h(x) + (1 - x) \ln (|\mathcal{X}| - 1).
$$
The first- and second-order derivatives of $g(x)$ are then given by
$$
g'(x) = \ln \frac{1 - x}{x(|\mathcal{X}| - 1)}
$$
and
$$
g''(x) = - \frac{1}{x(1 - x)}
$$
respectively. We then find that $g'(x) < 0$ for all $\frac{1}{|\mathcal{X}|} < x \le 1$, which implies $g(x_1) > g(x_2)$ for $\frac{1}{|\mathcal{X}|} \le x_1 < x_2 \le 1$. Hence we have
\begin{IEEEeqnarray*}{rCl}
H(P_{F_{n_l}(a^{n_l})}) &\le &g(P_{F_{n_l}(a^{n_l})}(a)) \\
&\ltvar{(a)} &g(\frac{1}{|\mathcal{X}|} + \epsilon_1) - g(\frac{1}{|\mathcal{X}|}) + g(\frac{1}{|\mathcal{X}|}) \\
&= &\ln|\mathcal{X}| - \epsilon_2
\end{IEEEeqnarray*}
for sufficiently large $l$, where $\epsilon_2 = g(\frac{1}{|\mathcal{X}|}) - g(\frac{1}{|\mathcal{X}|} + \epsilon_1) > 0$, and (a) follows from \eqref{eq:ProbabilityOfNonzeroSymbol} and the strictly monotonically decreasing property of $g$. This contradicts the inequality \eqref{eq:SystemticCode} and hence concludes the corollary.
\end{proofof}

\begin{proofof}{Corollary \ref{co:SparseMatrix}}
According to the condition \eqref{eq:MatrixDensityCondition}, there exits a positive real number $\epsilon_1$ and a subsequence $\{n_l\}_{l=1}^\infty$ such that
$$
D(A_{n_l \times m_{n_l}}) < 1 - \frac{1}{|\mathcal{X}|} - \epsilon_1
$$
for sufficiently large $l$. Furthermore, from the definition \eqref{eq:DefinitionOfMatrixDensity}, it follows that there exits a sequence $\{k_l\}_{l=1}^\infty$ such that
$$
\frac{1}{m_{n_l}} E[\sum_{j=1}^{m_{n_l}} 1\{A_{n_l \times m_{n_l}}(k_l, j) \ne 0\}] < 1 - \frac{1}{|\mathcal{X}|} - \epsilon_1
$$
for sufficiently large $l$. Let $1$ be the multiplicative identity in $\mathcal{X}$, and for each $l$, define a sequence $\seq{x}_l = 0^{k_l-1} 1 0^{n_l-k_l}$. Clearly, we have
\begin{equation}\label{eq:SparseMatrix1}
\lim_{l \to \infty} H(P_{\seq{x}_l}) = 0.
\end{equation}
Next, let $Y^{m_{n_l}} = F_{n_l}(\seq{x}_l)$, and then we find that
\begin{IEEEeqnarray*}{rCl}
E[P_{Y^{m_{n_l}}}(0)] &= &1 - \frac{1}{m_{n_l}} E[\sum_{j=1}^{m_{n_l}} 1\{A_{n_l \times m_{n_l}}(k_l, j) \ne 0\}] \\
&\ge &\frac{1}{|\mathcal{X}|} + \epsilon_1.
\end{IEEEeqnarray*}
Hence by similar arguments in the proof of Corollay \ref{co:SystematicCode}, we have
$$
E[H(P_{Y^{m_{n_l}}})] \levar{(a)} H(E[P_{Y^{m_{n_l}}}]) < \ln|\mathcal{X}| - \epsilon_2
$$
for some positive real number $\epsilon_2$ and sufficiently large $l$, where (a) follows from the fact that $H(P)$ is a concave function of $P$. This inequality together with \eqref{eq:SparseMatrix1} obviously contradicts Theorem \ref{th:DistanceProperty} with $\delta = 0$ if we assume that $\{F_n\}_{n=1}^\infty$ is asymptotically good, and hence concludes the corollary.
\end{proofof}

\section{Conclusion}\label{sec:Conclusions}

In this paper, we propose a lossless JSCC scheme based on linear codes and random interleavers for the transmission of correlated sources over MACs. To analyze the performance of this scheme, we establish an approach called ``code spectrum'' by the method of types. Then we show in Section \ref{subsec:IIIc} that linear codes with good joint spectra can be used to establish limit-approaching coding schemes for arbitrary correlated sources and MACs. Furthermore, in Section \ref{subsec:VariableRateCodingSchemes}, we further show that one good low-rate codes can be used to establish a limit-approaching coding scheme with multiple rates, and a novel idea called ``generalized puncturing'' is proposed.

A key concept introduced in this paper is that of ``pairwise independence of codewords''. Its sufficiency for JSCC is investigated in detail by Lemma \ref{le:FeinsteinLemmaForMAC} and Theorem \ref{th:GeneralizedDirectPartForMAC} in Section \ref{subsec:IIIb}. An important conclusion is thus drawn in Remark \ref{re:ArtOfJSCC} that a feasible approach to lossless JSCC is to generate a set of pairwise independent random codewords subject to a good conditional probability distribution. At the end of Section \ref{sec:JSCC}, we also list three important problems needed to be solved in future, namely, a problem of optimization, a problem of coding, and a problem of decoding. Undoubtedly, any essential advances on these problems will have vital significance on the design of JSCC.

Finally, based on our results and previous results \cite{JSCC:Bennatan200403, JSCC:Muramatsu200510, JSCC:Yang200503}, we can now list all the criteria and candidates of good linear codes for various coding problems of MACs, and all criteria are described in terms of spectrum requirements (see Table \ref{tab:CriteriaOfGoodLinearCodes}). It is pleasant that we can now answer the question ``What are good linear codes?'' for different coding problems in a unified framework established by our code-spectrum approach, and note that for the question ``What are good candidates for JSCC?'' we have found the answer, and readers may refer to \cite{JSCC:Yang200808} for details.
\begin{table*}[htbp]
\renewcommand{\arraystretch}{1.3}
\caption{Criteria and candidates of good linear codes for various coding problems of multiple-access channels}\label{tab:CriteriaOfGoodLinearCodes}
\centering
\begin{tabular}{p{1.8in}p{3in}p{1.5in}}
\hline
\begin{center}
\textbf{Coding problems}
\end{center}
&
\begin{center}
\textbf{Criteria of $\delta$-asymptotically good linear codes} \par
$\{F_n: \mathcal{X}^n \to \mathcal{Y}^{m_n}\}_{n=1}^\infty$
\end{center}
&
\begin{center}
\textbf{Good candidates}
\end{center}
\\
\hline
Slepian-Wolf coding (lossless source coding for noiseless MACs, \cite{JSCC:Slepian197307})
&
Kernel spectrum requirements (\cite{JSCC:Muramatsu200510, JSCC:Yang200503}):
$$
\limsup_{n \to \infty} \max_{P \in \mathcal{P}_n(\mathcal{X}) \backslash \{P_{0^n}\}} \frac{1}{n} \ln \frac{E[S_{\mathcal{X}}(\ker{F_n})(P)]}{S_{\mathcal{X}}(\mathcal{X}^n)(P)} \le \delta
$$
&
Pairty-check matrices of Turbo-like codes and LDPC codes (e.g., \cite{JSCC:Berrow199305, JSCC:Gallager196300, JSCC:Divsalar199809}). \\
Channel coding for MACs (e.g., \cite{JSCC:Cover199100})
&
Image spectrum requirements (implied by \cite{JSCC:Bennatan200403}):
$$
\limsup_{n \to \infty} \max_{Q \in \mathcal{P}_{m_n}(\mathcal{Y}) \backslash \{P_{0^{m_n}}\}} \frac{1}{m_n} \ln \frac{E[S_{\mathcal{Y}}(F_n(\mathcal{X}^n))(Q)]}{S_{\mathcal{Y}}(\mathcal{Y}^{m_n})(Q)} \le \delta
$$
&
Generator matrices of Turbo-like codes and LDPC codes.
\\
Lossless JSCC for MACs
&
Joint spectrum requirements:
$$
\limsup_{n \to \infty} \max_{\scriptstyle P \in \mathcal{P}_n(\mathcal{X}) \backslash \{P_{0^n}\}, \atop \scriptstyle Q \in \mathcal{P}_{m_n}(\mathcal{Y})} \frac{1}{n} \ln \frac{E[S_{\mathcal{X}\mathcal{Y}}(F_n)(P, Q)]}{S_{\mathcal{X}\mathcal{Y}}(\mathcal{X}^n \times \mathcal{Y}^{m_n})(P, Q)} \le \delta
$$
&
Linear codes with good joint spectra can be constructed based on sparse matrices. \cite{JSCC:Yang200808}
\\
\hline
\end{tabular}
\end{table*}

\section*{Acknowledgement}

The authors would like to thank Dr. T. Honold for his comments on the definition of linear codes with good joint spectra, which impel them to investigate the relation between a good random linear code and its sample codes in Proposition \ref{pr:RelationBetweenRLC&SLC}. The authors would also like to thank the anonymous reviewers for their helpful comments.

\bibliographystyle{IEEEtran} 
\bibliography{IEEEabrv,JSCC1}

\begin{thebibliography}{10}
\providecommand{\url}[1]{#1}
\csname url@samestyle\endcsname
\providecommand{\newblock}{\relax}
\providecommand{\bibinfo}[2]{#2}
\providecommand{\BIBentrySTDinterwordspacing}{\spaceskip=0pt\relax}
\providecommand{\BIBentryALTinterwordstretchfactor}{4}
\providecommand{\BIBentryALTinterwordspacing}{\spaceskip=\fontdimen2\font plus
\BIBentryALTinterwordstretchfactor\fontdimen3\font minus
  \fontdimen4\font\relax}
\providecommand{\BIBforeignlanguage}[2]{{%
\expandafter\ifx\csname l@#1\endcsname\relax
\typeout{** WARNING: IEEEtran.bst: No hyphenation pattern has been}%
\typeout{** loaded for the language `#1'. Using the pattern for}%
\typeout{** the default language instead.}%
\else
\language=\csname l@#1\endcsname
\fi
#2}}
\providecommand{\BIBdecl}{\relax}
\BIBdecl

\bibitem{JSCC:Cover198011}
T.~M. Cover, A.~E. Gamal, and M.~Salehi, ``Multiple access channels with
  arbitrarily correlated sources,'' \emph{{IEEE} Trans. Inf. Theory}, vol.~26,
  no.~6, pp. 648--657, Nov. 1980.

\bibitem{JSCC:Zhong200604}
Y.~Zhong, F.~Alajaji, and L.~L. Campbell, ``On the joint source-channel coding
  error exponent for discrete memoryless systems,'' \emph{{IEEE} Trans. Inf.
  Theory}, vol.~52, no.~4, pp. 1450--1468, Apr. 2006.

\bibitem{JSCC:Garcia200103}
J.~Garcia-Frias, ``Joint source-channel decoding of correlated sources over
  noisy channels,'' in \emph{Proc. Data Compression Conf.}, Mar. 2001, pp.
  283--292.

\bibitem{JSCC:Zhong200505}
W.~Zhong and J.~Garcia-Frias, ``{LDGM} codes for channel coding and joint
  source-channel coding of correlated sources,'' \emph{EURASIP Journal on
  Applied Signal Processing}, vol. 2005, no.~6, pp. 942--953, 2005.

\bibitem{JSCC:Zhao200611a}
Y.~Zhao and J.~Garcia-Frias, ``{Turbo} compression/joint source-channel coding
  of correlated binary sources with hidden {Markov} correlation,'' \emph{Signal
  Processing}, vol.~86, no.~11, pp. 3115--3122, 2006.

\bibitem{JSCC:Murugan200408}
A.~D. Murugan, P.~K. Gopala, and H.~E. Gamal, ``Correlated sources over
  wireless channels: Cooperative source-channel coding,'' \emph{{IEEE} J. Sel.
  Areas Commun.}, vol.~22, no.~6, pp. 988--998, Aug. 2004.

\bibitem{JSCC:Garcia200709}
J.~Garcia-Frias, Y.~Zhao, and W.~Zhong, ``{Turbo}-like codes for transmission
  of correlated sources over noisy channels,'' \emph{{IEEE} Signal Process.
  Mag.}, vol.~24, no.~5, pp. 58--66, Sep. 2007.

\bibitem{JSCC:Zhao200611b}
Y.~Zhao, W.~Zhong, and J.~Garcia-Frias, ``Transmission of correlated senders
  over a {Rayleigh} fading multiple access channel,'' \emph{Signal Processing},
  vol.~86, no.~11, pp. 3150--3159, 2006.

\bibitem{JSCC:Csiszar198100}
I.~Csisz\'{a}r and J.~K{\"{o}}rner, \emph{Information Theory: Coding Theorems
  for Discrete Memoryless Systems}.\hskip 1em plus 0.5em minus 0.4em\relax New
  York: Academic Press, 1981.

\bibitem{JSCC:Bennatan200403}
A.~Bennatan and D.~Burshtein, ``On the application of {LDPC} codes to arbitrary
  discrete-memoryless channels,'' \emph{{IEEE} Trans. Inf. Theory}, vol.~50,
  no.~3, pp. 417--437, Mar. 2004.

\bibitem{JSCC:Divsalar199809}
D.~Divsalar, H.~Jin, and R.~J. McEliece, ``Coding theorems for ``{Turbo}-like''
  codes,'' in \emph{36th Allerton Conf. on Communication, Control, and
  Computing}, Sep. 1998, pp. 201--210.

\bibitem{JSCC:Cover197503}
T.~M. Cover, ``A proof of the data compression theorem of {Slepian} and {Wolf}
  for ergodic sources,'' \emph{{IEEE} Trans. Inf. Theory}, vol.~21, no.~2, pp.
  226--228, Mar. 1975.

\bibitem{JSCC:Gallager196300}
R.~G. Gallager, \emph{Low-Density Parity-Check Codes}.\hskip 1em plus 0.5em
  minus 0.4em\relax Cambridge, MA: MIT Press, 1963.

\bibitem{JSCC:Csiszar198207}
I.~Csisz\'{a}r, ``Linear codes for sources and source networks: Error
  exponents, universal coding,'' \emph{{IEEE} Trans. Inf. Theory}, vol.~28,
  no.~4, pp. 585--592, Jul. 1982.

\bibitem{JSCC:Han200300}
T.~S. Han, \emph{Information-Spectrum Methods in Information Theory}.\hskip 1em
  plus 0.5em minus 0.4em\relax Berlin: Springer, 2003.

\bibitem{JSCC:Iwata200511}
K.~Iwata and Y.~Oohama, ``Information-spectrum characterization of
  multiple-access channels with correlated sources,'' \emph{IEICE Trans.
  Fundamentals}, vol. E88-A, no.~11, pp. 3196--3202, Nov. 2005.

\bibitem{JSCC:Gallager196800}
R.~G. Gallager, \emph{Information Theory and Reliable Communications}.\hskip
  1em plus 0.5em minus 0.4em\relax New York: Wiley, 1968.

\bibitem{JSCC:Delsarte198207}
P.~Delsarte and P.~Piret, ``Algebraic constructions of {Shannon} codes for
  regular channels,'' \emph{{IEEE} Trans. Inf. Theory}, vol.~28, no.~4, pp.
  593--599, Jul. 1982.

\bibitem{JSCC:Muramatsu200510}
J.~Muramatsu, T.~Uyematsu, and T.~Wadayama, ``Low-density parity-check matrices
  for coding of correlated sources,'' \emph{{IEEE} Trans. Inf. Theory},
  vol.~51, no.~10, pp. 3645--3654, Oct. 2005.

\bibitem{JSCC:Yang200503}
S.~Yang and P.~Qiu, ``On the performance of linear {Slepian-Wolf} codes for
  correlated stationary memoryless sources,'' in \emph{Proc. Data Compression
  Conf.}, Snowbird, UT, Mar. 2005, pp. 53--62.

\bibitem{JSCC:Yang200808}
S.~Yang, Y.~Chen, T.~Honold, Z.~Zhang, and P.~Qiu, ``Constructing linear codes
  with good joint spectra,'' in \emph{Proc. CHINACOM 2008}, Hangzhou, China,
  Aug. 2008, pp. 926--931, an extended version entitled ``Constructing linear
  codes with good spectra'' is now in preparation and to be submitted to IEEE
  Trans. Inf. Theory.

\bibitem{JSCC:Gilbert195210}
E.~N. Gilbert, ``A comparison of signaling alphabets,'' \emph{Bell Systems
  Technical Journal}, vol.~31, pp. 504--522, Oct. 1952.

\bibitem{JSCC:Huffman200300}
W.~C. Huffman and V.~Pless, \emph{Fundamentals of Error-Correcting
  Codes}.\hskip 1em plus 0.5em minus 0.4em\relax New York: Cambridge University
  Press, 2003.

\bibitem{JSCC:Yang200610A}
S.~Yang and P.~Qiu, ``On the performance of lossless joint source-channel
  coding based on linear codes,'' in \emph{Proc. Inf. Theory Workshop},
  Chengdu, China, Oct. 2006, pp. 160--164.

\bibitem{JSCC:Slepian197307}
D.~Slepian and J.~K. Wolf, ``Noiseless coding of correlated information
  sources,'' \emph{{IEEE} Trans. Inf. Theory}, vol.~19, no.~4, pp. 471--480,
  Jul. 1973.

\bibitem{JSCC:Berrow199305}
C.~Berrow, A.~Glavieux, and P.~Thitimajshima, ``Near {Shannon} limit
  error-correcting coding and decoding,'' in \emph{Proc. Int. Conf. Commun.},
  May 1993, pp. 1064--1070.

\bibitem{JSCC:Cover199100}
T.~M. Cover and J.~A. Thomas, \emph{Elements of Information Theory}.\hskip 1em
  plus 0.5em minus 0.4em\relax John Wiley \& Sons, 1991.

\end{thebibliography}

\begin{biographynophoto}{Shengtian Yang}
(S'05--M'06) was born in Hangzhou, Zhejiang, China, in 1976. He received the B.S. and M.S. degrees in biomedical engineering, and the Ph.D. degree in electrical engineering from Zhejiang University, Hangzhou, China in 1999, 2002, and 2005, respectively.

From June 2005 to December 2007, he was a Postdoctoral Fellow at the Department of Information Science and Electronic Engineering, Zhejiang University. Currently, he is an Associate Professor at the Department of Information Science and Electronic Engineering, Zhejiang University. His research interests include information theory and coding theory.
\end{biographynophoto}

\begin{biographynophoto}{Yan Chen}
(S'06) was born in Hangzhou, Zhejiang, China, in 1982. She received her B.Sc degree in information and communication engineering from Zhejiang University, Hangzhou, China, in 2004.

She is expected to receive her Ph.D degree in information and communication engineering from Zhejiang University in 2009. Since Jan 2008, She has been a visiting researcher in the Department of Electronic and Computer Engineering, Hong Kong University of Science and Technology, under the supervision of Prof. Vincent Lau. Her current research interests lie in combined information theory and queueing theory in wireless communications and networking, with particular emphasis on exploiting communication opportunities via user cooperation and cognition.
\end{biographynophoto}

\begin{biographynophoto}{Peiliang Qiu}
(M'03) was born in Shanghai, China, in 1944. He received the B.S. degree from the Harbin Institute of Technology, Harbin, China, in 1967 and the M.S. degree from the Graduate School of Chinese Academy of Science, Beijing, in 1981, respectively, both in electronics engineering.

During the period from 1968 to 1978, he worked in Jiangnan Electronic Technology Institute as an research engineer. Since November 1981, he has been with Zhejiang University, Hangzhou, China, where he is currently a Professor in the Department of Information Science and Electronic Engineering. His current research interests include digital communications, information theory, and wireless networks.
\end{biographynophoto}

\end{document}